\documentclass[onecolumn]{aa}
\usepackage{graphicx}
\usepackage{psfig}
\usepackage{txfonts}
\newfont{\myfont}{cmmib10}

\newcommand{\balpha}{\hbox{\myfont \symbol{11} }}
\newcommand{\bbeta}{\hbox{\myfont \symbol{12} }}

\newcommand{\bfeta}{\hbox{\myfont \symbol{17} }}
\newcommand{\btheta}{\hbox{\myfont \symbol{18} }}

\newcommand{\bzeta}{\hbox{\myfont \symbol{16} }}

\newcommand{\bxi}{\hbox{\myfont \symbol{24} }}

\newfont{\myfontsmall}{cmmib8}
\newcommand{\balphasm}{\hbox{\myfontsmall \symbol{11} }}

\def\lesssim{\mathrel{\hbox{\rlap{\hbox{\lower4pt\hbox{$\sim$}}}\hbox{$<$}}}}
\def\gtrsim{\mathrel{\hbox{\rlap{\hbox{\lower4pt\hbox{$\sim$}}}\hbox{$>$}}}}

\begin{document}
%
\title{Scattering of Gravitational Radiation}
\subtitle{Second Order Moments of the Wave Amplitude}
   \author{Jean-Pierre Macquart\inst{1}    }

   \offprints{Jean-Pierre Macquart}

   \institute{Kapteyn Astronomical Institute, University of Groningen, The Netherlands\\
              \email{jpm@astro.rug.nl}
             }

   \date{accepted 16/2/2004}

 \abstract{Gravitational radiation that propagates through an inhomogeneous mass distribution is subject to random gravitational lensing, or scattering, causing variations in the wave amplitude and temporal smearing of the signal.  A statistical theory is constructed to treat these effects. The statistical properties of the wave amplitude variations are a direct probe of the power spectrum of the mass distribution through which the waves propagate.  Scattering temporally smears any intensity variations intrinsic to a source emitting gravitational radiation, rendering variability on time scales shorter than the temporal smearing time scale unobservable, and potentially making the radiation much harder to detect.  Gravitational radiation must propagate out through the mass distribution of its host galaxy before it can be detected at the Earth.  Plausible models for the distribution of matter in an $L_*$ host galaxy suggest that the temporal smearing time scale is at least several milliseconds due to the gas content alone, and may be as large as a second if dark matter also scatters the radiation.  The smearing time due to scattering by any galaxy interposed along the line of sight is a factor $\sim 10^5$ times larger.  Gravitational scattering is an excellent probe of matter on parsec and sub-parsec scales, and has the potential to elucidate the nature of dark matter.
   }


   \maketitle
%
\section{Introduction}

Current and planned gravitational wave detectors have the potential to identify sources of gravitational radiation out to extragalactic distances and, for some types of sources, out to cosmological distances (Cutler \& Thorne 2002) over a wavelength range spanning more than 8 orders of magnitude.
Instruments such as the Laser Interferometer Gravitational Wave Observatory (LIGO) are sensitive to radiation in the range $10-10^4$\,Hz, while the planned Laser Interferometer Space Antenna (LISA) experiment will be sensitive to radiation from frequencies $10^{-1} -10^{-4}$\,Hz.  Other methods of detection, such as those due to pulsar timing observations, are sensitive to nano-Hertz gravitational radiation (Sazhin 1978, Detweiler 1979).  As the detection of gravitational radiation requires extremely sensitive apparatus, consideration has been devoted as to whether gravitational lensing by intervening massive objects might focus the radiation and enhance its detection likelihood (Ruffa 1999).

These lensing calculations (Ohanian 1974, Bliokh \& Minakov 1975, Bontz \& Haugan 1981, Thorne 1983, Deguchi \& Watson 1986a) are performed in the regime of physical optics.  
Diffractive effects are important because the wavelength of gravitational radiation is large compared to the Schwarzschild radius of any lensing object. For a wave of frequency $\nu$, diffractive effects are important for lens masses $\la 10^5 \nu^{-1} \,$M$_\odot$ (Takahashi \& Nakamura 2003).  Thus lensing by almost any intervening single object must be treated using physical optics.  Moreover, even for lens masses in which geometric optics is valid, it is preferable to work in the framework of physical optics because it allows one to compute certain quantities related to the wave amplitude that are in principle measurable with gravitational wave detectors but are not calculable under the framework of geometric optics.

Foregoing lensing treatments have largely been limited to the effects of a single massive object on the wave amplitude.  The approach has hitherto been from a deterministic rather than a statistical point of view, in that the mass profile of the lensing object is well specified.  However, a statistical approach is more appropriate under many circumstances.

The argument for a statistical approach stems from the large number of objects that are likely to contribute to the wave amplitude in many astrophysically interesting situations.  There are two senses in which a statistical approach is applicable to the gravitational lensing of gravitational radiation.  One is when a large number of lensing objects can drift in front of the line of sight to a source, so that an average over time would involve a sufficiently large number of objects to merit an average over an ensemble of lensing objects.  A second, more restrictive, sense applies when a large number of objects contributes to the observed wave amplitude at any instant.  This latter sense is likely to apply in practice.  The wave amplitude of a scattered signal is determined by the phase fluctuations caused by lensing objects in a region at least as large as the first Fresnel zone (see, e.g., Narayan 1992).  For gravitational lensing occurring at distance $D$ from the observer, the Fresnel radius $r_{\rm F}  \sim \sqrt{c D/2 \pi \nu}$, is of order a parsec for lensing of $\nu \sim 1\,$Hz radiation at cosmological distances.  Thus the area defined by the first Fresnel zone encompasses contributions from a large number of objects.  Consider, for example, the lensing influence of stars on the gravitational radiation passing through the spiral arm of a galaxy.  Even the volume of a line of sight orthogonal to the disk axis of a spiral galaxy with radius $r_{\rm F}\approx 1\,$pc would typically encompass many thousands of stars.  The argument for a statistical approach is strengthened when one considers the possible contribution of gas and of dark matter, both of which may be inhomogeneous on sub-parsec scales.


Scattering gives rise to a number of important effects that influence the detectability of signal.  The two that motivate the present treatment are (i) the focusing and defocusing of radiation by phase fluctuations, leading to temporal variability in the observed wave amplitude, and (ii) the temporal smearing of intrinsic intensity\footnote{Throughout the text we refer to the quantity defined by $\tilde \phi(\nu)$ in Eq. (\ref{Fresnel}) as the (complex) wave amplitude at frequency $\nu$, while the quantity $| \tilde \phi(\nu) |^2(t)$ is the intensity of the radiation at frequency $\nu$ and time $t$.  Thus intensity variations refer to changes in the quantity $|\tilde \phi (\nu) |(t)$ with time.} fluctuations in the source.  

Temporal fluctuations in the wave field occur whenever there is movement of the lensing system relative to the line of sight to the source of gravitational radiation.  Fluctuations in the gravitational potential drifting transverse to the line of sight induce both phase and amplitude fluctuations in the gravitational radiation, leading to temporal decorrelation of the wave field on sufficiently long time scales.  This effect is important if the decorrelation time scale associated with gravitational scattering is short compared to the time scale on which the wave field would oscillate in the absence of scattering.  In this paper we concentrate on fluctuations in the wave amplitude itself, while in a second paper we consider fluctuations in the intensity.

Temporal smearing is due to the fact that the radiation reaching an observer can arrive from a number of different angles.  This effect is significant when the phase fluctuations induced by the lensing are sufficiently large that radiation from outside the Fresnel radius is scattered toward the observer.  The radiation then arrives from a number of different regions in the scattering region, in an effect known as multipath propagation. There is a range of propagation times associated with the range of angles over which the radiation arrives. The effect is important because it potentially smears out the intrinsic intensity variations of a source.  For instance, time variations of the source would be undetectable if the intrinsic variability time scale were significantly shorter than the scatter broadening time.

The scattering of gravitational radiation potentially constitutes an exceptional probe of the distribution of matter out to the distances to which sources can be detected.  In this context it is useful to compare gravitational scattering to the interstellar scattering (ISS) of radio wavelength radiation by the diffuse ionized component of our Galaxy's interstellar medium.  Both scattering phenomena require a statistical approach and occur in a regime in which physical optics is important.  ISS yields information on the power spectrum of turbulent fluctuations in the ionized plasma distribution of our own Galaxy over a range of scales spanning some five orders of magnitude in wavenumber (Armstrong, Rickett \& Spangler 1995).  This information is gleaned primarily from pulsars which are effectively point sources, and thus constitute excellent probes of fluctuations in the interstellar medium.  Other radio sources are less subject to ISS, and their larger angular diameters severely truncate the range of wavenumbers over which the power spectrum can be measured.  Gravitational wave sources are the analogues of pulsars in the regime of gravitational scattering. They emit coherently and possess small angular diameters so the source structure does not influence the scattering characteristics, unlike most sources of electromagnetic radiation that are subject to gravitational lensing.  Thus the scattering of gravitational radiation is a fine-scale probe of the distribution of dark matter in the local universe.

The close relationship between gravitational and interstellar scattering implies that many phenomena identified in this paper are analogous to effects previously identified in the context of interstellar scintillation.  For example, the temporal broadening of gravitational radiation is analogous to the temporal broadening of pulsar radiation, and it poses similar limitations on source detectability (see the review by Rickett 1977 and references therein).
  
The outline of the paper is as follows.  In Sect.\,2 we review the effect of an arbitrary mass distribution on the amplitude of a gravitational wave.  In  Sect.\,3 we compute the temporal variability of the wave amplitude by considering the power spectrum of the wave amplitude fluctuations.  The effect of temporal broadening on the scattered signal is considered in Sect.\,4.  Estimates of the power spectrum of mass fluctuations due to the various constituents of a galaxy are presented in Sect.\,5.  In Sect.\,6 we estimate the magnitude of scattering effects and discuss their relevance to the observability of gravitational waves.  The conclusions are presented in Sect.\,7.

\section{Propagation of the wave amplitude}

We briefly outline the equations governing the propagation of gravitational radiation in the gravitational potential of a distribution of lensing objects.   One writes the spacetime metric in the form
\begin{eqnarray}
ds^2 = - (1-2 U({\bf r})) c^2 dt^2 + (1-2U({\bf r})) d{\bf r}^2 \equiv g_{\mu \nu}^{(B)} dx^\mu dx^\nu, \label{metric}
\end{eqnarray}
where $U({\bf r}) \ll 1$ is the gravitational potential of the lensing objects.  Consider a linear perturbation $h_{\mu \nu}$ in the background metric tensor, $g_{\mu \nu}^{(B)}$: $g_{\mu \nu} = g_{\mu \nu} + h_{\mu \nu}$.  If the wavelength of the gravitational radiation is much smaller than the typical radius of curvature of the background metric, then one has
$
2 R_{\alpha \mu \beta \nu}^{(B)} h^{\alpha \beta} = 0,
$ where $R_{\alpha \mu \beta \nu}^{(B)}$ is the background Riemann tensor, and one identifies the gravitational wave as 
\begin{eqnarray}
h_{\mu \nu} = \phi \, e_{\mu \nu},
\end{eqnarray}
where $e_{\mu \nu}$ is the polarization tensor of the gravitational wave ($e^\mu_\mu=0$, $e_{\mu \nu} e^{\mu \nu} = 2$) and $\phi=\phi(t,{\bf r})$ is the scalar wave amplitude.  This scalar wave propagates according to 
$
\partial_\mu (\sqrt{ -g^{(B)} } g^{ (B)\mu\nu} \partial_\nu \phi) = 0,
$ which, when combined with equation \,(\ref{metric}), takes the following simple form
\begin{eqnarray}
(\nabla^2 + \omega^2 ) \tilde{\phi}  = 4 \omega^2 U \tilde{\phi}, \label{WaveEqun}
\end{eqnarray}
where $\omega=2 \pi \nu$ and $\tilde{\phi}(\nu,{\bf r})$ is the temporal Fourier transform of the scalar wave amplitude.  


We consider the solution of Eq. (\ref{WaveEqun}) with reference to the lensing geometry shown in Fig.\,1.
A point source of unit intensity is located an angular diameter distance $D_{\rm S}$ from the observer and distance $D_{\rm LS}$ from the lens plane.  The lens plane is located an angular diameter distance $D_{\rm L}$ from the observer's plane, and the observer is at location ${\bf X}'$ on this plane.  
The gravitational perturbations are assumed to be located on a thin lensing plane. 
One might be concerned that the thin lensing approximation used here, while applicable to the lensing of electromagnetic radiation, is not applicable to gravitational radiation because of its short wavelength.  The depths of most lensing objects are small compared to the wavelength of gravitational radiation. However, the thin lens approximation is still valid because it depends only on the weakness of the wave amplitude and not on the short wavelength assumption (Thorne 1983).

The wave amplitude has the solution (Schneider, Ehlers \& Falco 1992)
\begin{eqnarray}
\tilde{\phi}(\nu,{\bf X}') = \frac{D_S \nu  (1+z_L) }{c D_L D_{LS}} e^{-i \pi/2} \int d^2{\bf x} \exp[2\pi i  \nu t_d({\bf x},{\bf X}')], \label{Fresnel}
\end{eqnarray}
where the time delay is, apart from a constant, given by
\begin{eqnarray}
t_d ({\bf x},{\bf X}') = \frac{D_S (1+z_L)}{c D_L D_{LS} } \left[\frac{1}{2} \left( {\bf x} - \frac{D_{LS}}{D_{S}} {\bf X}' \right)^2 -\Psi({\bf x})  \right].
\end{eqnarray}
The term $D_{LS}/D_{S}$ in front of the observer's plane co-ordinate ${\bf X}'$ is a correction factor that accounts for the spherical nature of the wavefront in the mapping of lens plane co-ordinates to co-ordinates on the observer's plane (see Goodman \& Narayan 1989).  When the source is located sufficiently far behind the lens plane the incident wavefront may be regarded as planar and this correction factor tends to unity. We introduce the scaled co-ordinate ${\bf X}={\bf X}' D_{LS}/D_{S}$ to retain the symmetry between lens and observer plane co-ordinates below.

The phase delay associated with the surface mass distribution, $\Sigma({\bf x})$, is
\begin{eqnarray}
\Psi ({\bf x}) &=& \frac{ 4 G D_L D_{LS}}{D_S c^2} \int d^2{\bf x}' \Sigma({\bf x}') \ln \left(\frac{| {\bf x}-{\bf x}'|}{ x_0} \right),
 \end{eqnarray}
where $x_0$ is an arbitrary normalizing constant that is of no interest because it only introduces an arbitrary constant phase delay; we henceforth write $x_0=1$ to set this phase delay to zero.  

It is convenient to recast the Fresnel-Kirchoff integral (\ref{Fresnel}) in the form 
\begin{eqnarray}
\tilde{\phi}(\nu,{\bf X}) =  \frac{e^{-i \pi/2} }{2 \pi r_{\rm F}^2} \int d^2 {\bf x} \exp \left[ \frac{i}{2 r_{\rm F}^2} \left( {\bf x}- {\bf X} \right)^2 +   i K  \int  d^2{\bf x}' \Sigma({\bf x}') \ln \left(\frac{| {\bf x}-{\bf x}'|}{{x}_0} \right)  \right], 
\qquad \qquad \hbox{where } K = -8 \pi \frac{1+z_L}{ \lambda}   \frac{G}{c^2},
\label{uGrav}
\end{eqnarray}
and where we define the curved-spacetime generalization of the Fresnel scale in Euclidean space as
\begin{eqnarray}
r_{\rm F}^2 = \frac{D_L D_{LS} \lambda }{2 \pi D_S (1 + z_L)}, \label{rF}
\end{eqnarray}
and the phase delay due to the gravitational potential is identified explicitly as 
\begin{eqnarray}
\psi ({\bf x}) = K \int d^2{\bf x}' \, \Sigma({\bf x}') \ln \left( \frac{|{\bf x}-{\bf x}'|}{x_0}\right).
\end{eqnarray}
The approach adopted in this paper involves regarding the gravitational phase delay, $\psi$, as a random variable.  We define $r_{\rm diff}$ as the length scale on the lens plane over which the root-mean-square phase delay changes by one radian.  The magnitude of effects due to gravitational scattering are quantized by comparing the length scale of gravitational phase fluctuations, $r_{\rm diff}$, to the length scale of geometric phase fluctuations, characterized by the Fresnel scale, $r_{\rm F}$.  Scattering effects are important when $r_{\rm diff}$ is comparable to or smaller than $r_{\rm F}$.  To see this, consider the generic properties of the wave field as determined by Eq. (\ref{uGrav}).
The dominant contributions to the wave field come from regions where the argument of the exponential in Eq. (\ref{uGrav}) varies slowly as a function of ${\bf x}$.  For small gravitational phase fluctuations on the length scale $r_{\rm F}$ (i.e. $r_{\rm diff} \gg r_{\rm F}$), only the region around $|{\bf x}-{\bf X}| < r_{\rm F}$ contributes to the wave field, and mild phase curvature on the scale of $r_{\rm F}$ gives rise to mild focusing and defocusing of the gravitational wave.  For very large phase fluctuations (i.e. $r_{\rm diff} \ll r_{\rm F}$) many separate regions on the lensing plane contribute to the observed wave amplitude.  This is because there are many regions on the lens plane for which the exponential term in (\ref{uGrav}) varies slowly as a function of ${\bf x}$, since the large phase fluctuations can offset the contribution of the $({\bf x}-{\bf X})^2/ 2 r_{\rm F}^2$ term. 


\begin{figure}
\centerline{\psfig{file=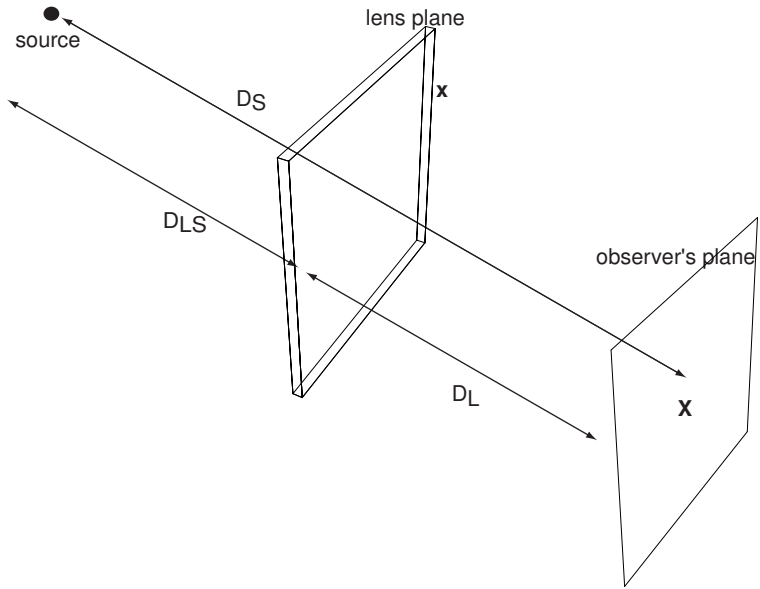,width=10cm}}
\caption{The lensing geometry.  Co-ordinates on the lens plane are denoted with ${\bf x}$ and those on the observer's plane by ${\bf X}$.}
\end{figure}

\subsection{The effect of finite source size}
As sources of gravitational radiation are compact and spatially coherent it suffices to treat the radiation from these systems as point-like (Mandzhos 1981, Ohanian 1983, Schneider \& Schmid-Burgk 1985, Deguchi \& Watson 1986b, Peterson \& Falk 1991).  Thus equation (\ref{uGrav}) suffices as a practical  description of the wave amplitude observed from any source of gravitational radiation.  Equation (\ref{uGrav}) also describes the gravitational lensing of electromagnetic radiation from compact objects, but it needs to be modified once the source angular diameter becomes sufficiently large.  In practice, most sources of electromagnetic radiation are not spatially coherent and are sufficiently large that effects due to their finite angular diameter are important.

The effect of finite source size on the temporal variability of gravitationally scattered radiation is not discussed in the present paper, but the results in this paper are relevant to the lensing of electromagnetic radiation when the source size does not exceed the characteristic angle through which scattering deflects the radiation.  Stated another way, source size effects are important when the source angular diameter exceeds the angular scale over which the wave field of a point source would vary due to inhomogeneities in the scattering medium (e.g. Little \& Hewish 1966, Salpeter 1967).
To see this, one can attribute wave field decorrelation length scales to both the effects of source size and of scattering.  Define the length scale on the observer's plane, $X_{\rm src}$, over which structure in the source causes the wave field to decorrelate (i.e. $\langle \tilde \phi({\bf X}_{\rm src}+{\bf x}') \tilde \phi^*({\bf x}') \rangle$ declines on the length scale $X_{\rm src}$).  For an incoherent source of angular size $\theta_{\rm src}$, this length scale is $X_{\rm src} = \lambda/2 \pi \theta_{\rm src}$.  One can define a similar length scale $X_{\rm scat}=\lambda / 2 \pi \theta_{\rm scat}$ over which the wave field from a point source would decorrelate due to scattering.  Source size effects are therefore unimportant when the source decorrelation scale exceeds the scattering decorrelation scale, $X_{\rm src} \ga X_{\rm scat}$ or, equivalently, when the angular scale of the scattering pattern from a point source exceeds the source angular diameter, $\theta_{\rm scat} \ga \theta_{\rm src}$.

\subsection{Statistics of the mass fluctuations}
In the following discussion we are interested in the statistical properties of the wave amplitude as influenced by gravitational lensing.  We explicitly ignore (non-stochastic) large-scale phase gradients due to the overall mass distribution in a lensing system.  Large scale phase gradients contribute to macrolensing, but are not relevant to the effects under consideration here.

Since the statistics of wave amplitude fluctuations depend on the statistical properties of the phase, $\psi({\bf x})$, we must characterize the fluctuations in the mass surface density, $\Sigma({\bf x})$, that drive the phase perturbations on the lensing plane.   The two most useful quantities are the mass surface density autocorrelation function
\begin{eqnarray}
C_{\Sigma} ({\bf r}) =  \langle  \Delta \Sigma({\bf r}'+{\bf r})  \Delta \Sigma({\bf r}') \rangle = \langle [ \Sigma({\bf r}'+{\bf r}) - \bar{\Sigma} ]  [ \Sigma({\bf r}') - \bar{\Sigma} ] \rangle,
\end{eqnarray}
and its associated quantity, the mass surface density structure function,
\begin{eqnarray}
D_{\Sigma} ({\bf r}) = \langle [\Sigma ({\bf r}'+{\bf r}) - \Sigma ({\bf r}')]^2  \rangle = 2 [C_{\Sigma}(0) - C_{\Sigma}({\bf r})],
\end{eqnarray}
which describes the mean square difference in the mass surface density between two points separated by a vector ${\bf r}$ on the lensing plane.  The autocorrelation function is related directly to the power spectrum of mass surface density fluctuations $\Phi_\Sigma(q_x,q_y)$ as follows:
\begin{eqnarray}
C_{\Sigma} ({\bf r}) = \int \frac{d^2{\bf q}}{(2 \pi)^2 } \, \Phi_\Sigma({\bf q}) \, e^{i {\bf q} \cdot {\bf r}}.
\end{eqnarray}
These average quantities are written as functions of separation ${\bf r}$ only, because it is assumed that the statistical properties of $\Sigma$ do not depend on the position where they are measured, which is to say that the statistical properties of $\Sigma$ are wide-sense stationary (see, e.g., Mandel \& Wolf 1995).
Although this assumption is commonly used in scattering theory, it is obviously flawed to some degree because the statistical properties of the lensing matter are expected to vary as, for instance, one moves from a line of sight intersecting the centre of a galaxy towards one that intersects its rim.  However, it is an excellent approximation in practice since we are concerned with the statistical properties of the matter distribution mainly on scales less than $\sim 10\,$pc.

It is often convenient to relate correlations in the gravitational phase $\psi({\bf x})$ to the power spectrum of gravitational mass fluctuations. The autocorrelation in phase fluctuations $\psi({\bf x})$ and $\psi({\bf x}+{\bf r})$ is derived using the identity, 
\begin{eqnarray}
\int d^2 \balpha \ln |\balpha | e^{i {\bf q} \cdot \balphasm} &=& - 2 \pi q^{-2}, \label{logFT}
\end{eqnarray} 
which yields the following relation between phase autocorrelation and the mass power spectrum,
\begin{eqnarray}
\langle \Delta \psi({\bf x}) \Delta \psi({\bf x}+{\bf r}) \rangle = K^2 \int d^2 {\bxi} d^2 {\bzeta} \ln | \bxi | \,\ln | \bzeta |  \, C_\Sigma ({\bxi-\bzeta-{\bf r} }) 
=   K^2 \int d^2{\bf q} \, e^{i {\bf q} \cdot {\bf r}} \,q^{-4} \,  \Phi_\Sigma ({\bf q}). \label{CtoPhi}
\end{eqnarray}
The structure function of phase fluctuations $D_\psi({\bf r})$ is related to the mass power spectrum
according to
\begin{eqnarray}
D_\psi({\bf r}) = 2 \,K^2 \int d^2{\bf q} \, \left[ 1-e^{i {\bf q} \cdot {\bf r}} \right] \,q^{-4} \,  \Phi_\Sigma ({\bf q}). \label{DtoPhi}
\end{eqnarray}
These relations simplify the calculations performed below and allow us to express the results in physically useful quantities.

\section{Wave amplitude fluctuations}

Given the exact mass distribution on the lens plane, eq. (\ref{uGrav}) can be evaluated directly to find the wave amplitude on the observer's plane.  The mass distribution on the lensing plane is usually unknown, but it is possible to express the statistical moments of the wave amplitude in terms of statistical properties of the mass distribution.
In the following subsection we evaluate the variability due to gravitational scattering by computing the covariance of wave amplitude fluctuations across the observer's plane.

\subsection{Mean visibility} \label{MeanVis}

We wish to compute the time scale on which the wave amplitude of a source lensed by random mass inhomogeneities fluctuates.  The temporal fluctuations are characterized by the covariance between the wave amplitude received at a frequency $\nu$ and time $t$ with those detected at the same frequency at some later time $t+\Delta t$: $\langle \tilde\phi(\nu;t) \tilde \phi^*(\nu;t+\Delta t) \rangle$.  The angular brackets here refer to an average over all possible statistical ensembles of lensing objects which, here, is equivalent to an average over all times $t$.   

The mass fluctuations on the lensing plane are assumed to be fixed, and the lensing plane advects past an observer at some effective velocity ${\bf v}_{\rm eff}$.  This is known as the frozen screen approximation.  A stationary observer who samples the wave field at position ${\bf X}$ at times $t_1,t_2,\ldots$ would observe exactly the same wave field as an observer who moves across the observing plane, measuring the wave fields at all the positions ${\bf X}+{\bf v}_{\rm eff} t_1$, ${\bf X}+{\bf v}_{\rm eff} t_2$, \ldots on the observing plane.  An average over time is equivalent to an average over all positions ${\bf X}$ on the observer's plane.  To be specific, the temporal covariance of the wave field is equivalent to the covariance measured between two separated receivers located at positions ${\bf x}$ and  ${\bf x} + {\bf X} ={\bf x}+ {\bf v}_{\rm eff} \Delta t$.  
The specific expression for the effective velocity, in terms of the velocities of the source, lensing plane and observer is (see Appendix \ref{ScatSpeed})
\begin{eqnarray}
{\bf v}_{\rm eff} =  {\bf v}_{\rm screen} - {\bf v}_{\rm Earth} \left(1 -\frac{D_L}{D_S} \right) - {\bf v}_{\rm src} \left( \frac{D_S}{D_L} \right) .
\end{eqnarray}

To compute the wave field covariance we consider the wave amplitude measured by two detectors a distance ${\bf X}$ apart on the observing plane: one receiver is positioned at $-{\bf X}/2$ while the other is located at ${\bf X}/2$.  In light of its connection with the corresponding quantity in interferometry (see Thomson, Moran \& Svenson 1986), we refer to this covariance as the mean visibility, or mutual coherence, of the gravitational wave field.
Using the the Fresnel integral derived in eq. (\ref{uGrav}), the mutual coherence is 
\begin{eqnarray}
V(\nu;{\bf X}) \equiv \langle \tilde \phi (\nu;-{\bf X}/2) \tilde \phi^* (\nu;{\bf X}/2) \rangle &=& \frac{1}{(2 \pi r_{\rm F}^2)^2} \int d^2 {\bf x} d^2 {\bf x}' \exp \left\{ \frac{i}{2 r_{\rm F}^2} \left[ \left( {\bf x} + {\bf X}/2 \right)^2 - \left( {\bf x}' - {\bf X} /2 \right)^2 \right] \right.
\nonumber \\  &\null&
\left. \hskip 2in + i K \int d^2{\bxi} \, \Sigma ({\bxi}) 
\left[ \ln | {\bxi} - {\bf x} | - \ln | {\bxi} - {\bf x}' | \right] \right\}.
\end{eqnarray}
It is useful to make the change of variable ${\bf r}={\bf x}- {\bf x}'$ and ${\bf s} = ({\bf x}+{\bf x}')/2$, so the visibility simplifies to
\begin{eqnarray}
V(\nu;{\bf X}) &=&  \frac{1}{(2 \pi r_{\rm F}^2)^2} \int d^2 {\bf r} d^2 {\bf s} \exp \left[ \frac{i {\bf s} \cdot ({\bf r} + {\bf X})}{r_{\rm F}^2 } + i K  \int d^2{\bxi}\,  \ln | {\bxi} |\  \left[  \Sigma ({\bxi}+{\bf r}/2+{\bf s}) - 
\Sigma(\bxi - {\bf r}/2 +{\bf s}) \right] \right]. \label{VisRaw}
\end{eqnarray}
We assume that the mass-induced phase fluctuations are wide-sense stationary, so that the difference of the mass surface density 
$\Sigma ({\bxi}+{\bf r}/2+{\bf s}) - \Sigma(\bxi - {\bf r}/2 +{\bf s})$ depends only on the separation ${\bf r}$. The integral over ${\bf s}$ then yields a factor $(2 \pi r_{\rm F}^2)^2 \delta^2 ({\bf r}+{\bf X})$ and after performing the remaining integral over ${\bf r}$, the visibility reduces to 
\begin{eqnarray}
V(\nu;{\bf X}) &=& \left\langle \exp \left[ i K \int d^2\bxi \, \Sigma (\bxi) \left(\ln |\bxi + {\bf X}/2| - \ln |\bxi - {\bf X}/2 \right) \right] \right\rangle. \label{VisRefined}
\end{eqnarray}
The average over the mass fluctuations is evaluated using the result
\begin{eqnarray}
\langle \exp[ i \alpha ] \rangle = e^{i \langle \alpha \rangle} \left[ 1 + i \langle \delta \alpha \rangle - \frac{1}{2} \langle \delta \alpha^2 \rangle - \frac{i}{3!} \langle \delta \alpha^3 \rangle + \frac{1}{4!} \langle \delta \alpha^4 \rangle - \ldots \right], \label{AvgExpansion}
\end{eqnarray}
for a random variable $\alpha$ with average $\langle \alpha \rangle$ and fluctuations $\delta \alpha$.  This approximation makes no specific assumptions about the distribution of the random variable $\delta \alpha$.
When the fluctuations in the gravitational phase, $\psi$, are small, it is sufficient to consider only the first three terms in this expansion:
\begin{eqnarray}
\langle V(\nu;{\bf X}) \rangle &\approx&  1 + i  K \langle Y \rangle - \frac{1}{2} K^2 \langle Y^2 \rangle, \label{MeanVisDefn} \qquad \hbox{ where } Y =   \int d^2\bxi \, \ln |\bxi | \,
\left[ \Sigma (\bxi-{\bf X}/2) - \Sigma (\bxi+{\bf X}/2) \right] .
\end{eqnarray}
The imaginary component of the contribution is zero because $\langle Y \rangle =0$; for mass  fluctuations which are wide-sense stationary (i.e. their statistical properties are independent of position) the average $\langle \Sigma (\bxi - {\bf X}/2) \rangle$  is equal to $\langle \Sigma (\bxi + {\bf X}/2) \rangle$.  The final term involving terms ${\cal O}(K^2)$ simplifies to 
\begin{eqnarray}
\langle Y^2 \rangle &=&   \int_{| \balpha | > 0 , | \bbeta | > 0} d^2 \balpha \, d^2 \bbeta \, \ln |\balpha| \, \ln |\bbeta | \left[\frac{ D_\Sigma(\balpha-\bbeta - {\bf X})}{2} + 
\frac{D_\Sigma(\balpha-\bbeta - {\bf X})}{2}  -  D_\Sigma(\balpha-\bbeta)\right], 
\end{eqnarray}
which, re-expressed in terms of the power spectrum of mass surface density fluctuations using equation (\ref{CtoPhi}), yields
\begin{eqnarray}
\langle V(\nu,{\bf X}) \rangle \approx 1 -  K^2 \int d^2 {\bf q} \, q^{-4} \, \Phi_\Sigma ({\bf q})  \left[ 1 - \cos ({\bf q} \cdot {\bf X}) \right]. \, \label{YsqrResult}
\end{eqnarray}

One obtains a clearer result on the length scale over which the scattered wave field varies by examining the power spectrum of wave amplitude fluctuations across the observer's plane, which is just the Fourier transform of the visibility:
\begin{eqnarray}
\langle \Phi_\phi ({\bf q}) \rangle \approx \int d^2{\bf X} \, e^{-i {\bf q} \cdot {\bf X}} \, \langle V(\nu;{\bf X}) \rangle &=& 
K^2 q^{-4} \Phi_\Sigma ({\bf q}) + (2 \pi)^2 \delta^2 ({\bf q}) \left[ 1 - K^2 \int d^2{\bf q}' \,  q'^{-4} \Phi_\Sigma ({\bf q}') \right].  \label{PhiResult1}
\end{eqnarray}
Apart from the unimportant zero-frequency ($q=0$) contribution, the spatial Fourier transform of the mean visibility yields a {\it direct} measurement of the power spectrum of mass density fluctuations weighted by  $q^{-4}$, which weights the measurement to the largest lensing structures present in the medium.  Measurements of the mean visibility potentially provide an elegant probe of the mass density fluctuations along the line of sight to a source of gravitational radiation.  

\subsubsection{Strong phase fluctuations}

The foregoing result embodied in equation (\ref{PhiResult1}) is only correct in the limit in which the phase fluctuations are small.  Using equation (\ref{DtoPhi}), we can express the visibility in terms of the phase structure function:  $\langle V(\nu,{\bf X}) \rangle \approx 1 -  D_\psi({\bf r})/2$.  This shows that the small phase approximation is only valid for sufficiently small baselines ${\bf r}$ such that the mean square difference in the phase delay due to random mass fluctuations across the lensing plane is less than one radian.   
 
It is possible to employ an alternative approximation to derive the mean visibility in a manner that is independent of the magnitude of the phase fluctuations.  However, this requires that we make specific assumptions about the statistical properties of the gravitational phase, $\psi$.  In particular, we assume that phase fluctuations are normally distributed, which allows the average over phase fluctuations to be performed using the result
\begin{eqnarray}
\langle \exp [i \psi ] \rangle = \exp[ i \langle \psi \rangle - \langle \delta \psi^2 \rangle /2].
\end{eqnarray}
Applying this result to the phase fluctuations in (\ref{VisRefined}), one derives an exact result for the mean visibility under this somewhat more restrictive assumption:
\begin{eqnarray}
V(\nu;{\bf X}) &=& \exp \left[ - \frac{D_\psi ({\bf X})}{2}\right] 
=  \exp \left[ - K^2 \int d^2 {\bf q} \, q^{-4} \left[ 1- e^{i {\bf q} \cdot {\bf X}}\right] \Phi_\Sigma({\bf q})  \right]. \label{VisGeneral}
\end{eqnarray}
This generalization of (\ref{YsqrResult}) shows that the wave amplitude on the observer's plane is only correlated over distances on which the mean square phase difference on the lens plane is less than one radian.  In order words, the wave field is only correlated over a distance $\sim r_{\rm diff}$. (Recall that $r_{\rm diff}$ is defined such that $D_\psi(r_{\rm diff}) =1$.)  The frozen screen approximation described above then implies that temporal fluctuations in the wave field occur on time scales as short as $t_{\rm var} \approx r_{\rm diff}/v_{\rm eff}$.  The covariance between wave amplitudes measured on time scales longer than $t_{\rm var}$ decreases exponentially quickly.

We can exploit the similarity between scattering due to gravitational mass perturbations and interstellar scattering in interpreting the physics embodied in equation (\ref{VisGeneral}).  For phase fluctuations $\phi({\bf x})$ in the interstellar medium, the visibility of a scattered point source is $V_{\rm ISS}(\nu,{\bf X}) = \exp [- D_\phi({\bf X})/2]$, which is identical in form to equation (\ref{VisGeneral}). The visibility is related to the apparent source brightness distribution (i.e. the source image), $I(\btheta)$, according to (e.g. Thomson et al. 1986) 
\begin{eqnarray}
I(\btheta) = \int d{\bf X} \,e^{i k {\bf X} \cdot \btheta} V(\nu,{\bf X}).
\end{eqnarray}
Since the \emph{average} visibility of a scattered source declines on a length scale $X=r_{\rm diff}$, the \emph{average} angular brightness distribution of the scattered source declines on an angular scale $\theta = 1/k r_{\rm diff}$.  The implication for observations of gravitational radiation is that a detector observing a scattered source receives radiation from a range of angles out to an  average angular-broadening scale $\theta_{\rm scat} = 1/k r_{\rm diff}$.  

Under certain circumstances it is appropriate to consider the variance in the visibility as a measure of the extent to which it deviates from its mean value.  The fourth moment of the wave field, which describes the covariance of fluctuations in the visibility, is computed in a subsequent paper (Macquart, in prep.). 

\section{Temporal broadening of the scattered signal}
Propagation effects can cause temporal smearing of gravitational radiation, leading to possible misinterpretation or even non-detection of  bursty phenomena.  Temporal broadening is important when a signal propagating through a random medium can potentially reach an observer along several different paths (this is known as multipath propagation).  As different time delays are associated with the ray paths, power from an impulsive event is smeared out in a process known as temporal broadening.    This can redistribute power in the emitted signal so that the temporally varying signal is smeared to a level below the detection threshold (see Fig.\,2).  

The temporal smearing of gravitational radiation is analogous to the temporal smearing of radio wavelength radiation received from pulsars.  For pulsar radiation this effect is due to scattering off electron density inhomogeneities in the interstellar medium, and is important for most pulsars located in the Galactic plane, as it limits the temporal resolution with which the pulse shape can be measured.  The effect severely limits the detectability of pulsar radiation when the temporal smearing time scale is comparable to the pulse period.  This is particularly important for the heavily scattered pulsars located near the Galactic centre, as the broadening time scale is so large that pulsar emission appears continuous rather than pulsed;  although the emission is still detected, it is not time-variable.

\begin{figure}
\centerline{\psfig{file=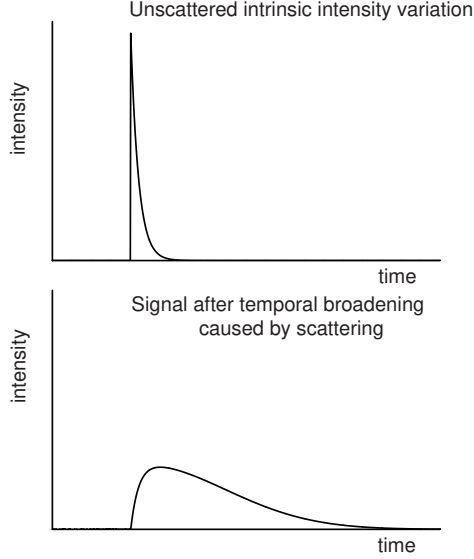,width=6cm}}
\caption{Temporal smearing of a time variable signal (i.e. one for which $|\tilde \phi(\nu)|(t)$ varies with time) smooths out the variations so that the radiation arrives at the detector over a longer interval and the source potentially falls below the detection threshold.}
\end{figure}

The treatment developed for the smearing of gravitational radiation is analogous to that derived for the temporal broadening of pulsar radiation due to interstellar scattering (Lee \& Jokipii 1975).  The main differences here are that perturbations in the gravitational potential drive the phase fluctuations rather than fluctuations in the electron density, and that the gravitational phase fluctuations scale like $\lambda^{-1}$, whereas those in an interstellar plasma are proportional to $\lambda$.

Before discussing effects caused by inhomogeneities in the gravitational potential, we first consider effects related to the propagation of the wave through a homogeneous gravitational field and its subsequent detection.  We explicitly decompose the wave function into its temporal Fourier components, $\tilde\phi(\omega)$ as follows:
\begin{eqnarray}
\phi(t) = \frac{1}{2 \pi} \int_{-\infty}^{\infty} \tilde \phi(\omega) e^{-i \omega t} \,d\omega. \label{WaveDecomp}
\end{eqnarray}
The first effect that limits the measurement of the wave function is the spectral response, or bandpass, of the detector, which we idealize as
\begin{eqnarray}
f_B(\omega) = \exp \left[ \frac{-(\omega-\omega_0)^2}{2 \Delta^2} \right],
\end{eqnarray}
where $\omega_0$ is the central frequency and $\Delta$ is the bandwidth.  The detected signal is then
\begin{eqnarray}
\tilde \phi_{\rm det}(t) = \frac{1}{2 \pi} \int_{-\infty}^{\infty} \tilde \phi(\omega)\, f_B(\omega) \, e^{-i \omega t} \, d\omega.
\end{eqnarray}
Consider the response to radiation with a sharp pulse at time $t=0$ propagating through free space.  Then $\phi(t)=\delta(z/c-t)$, $\tilde \phi(\omega) = e^{i k z}$ and the measured wave function is
\begin{eqnarray}
|\tilde \phi_{\rm det}(t)| = \frac{\Delta}{\sqrt{2 \pi}} \exp \left[- \frac{(t-z/c)^2 \Delta^2}{2} \right].
\label{FreeSpace}
\end{eqnarray}

Now we introduce effects related to propagation through a homogeneous gravitational field.  Consider first a homogeneous gravitational field which gives rise to a constant phase delay $\langle \psi \rangle$.  One can relate this to a decrease in the group velocity of the radiation with respect to $c$.  A phase delay $\psi$ gives rise to a time delay $\psi/\omega$ and, for propagation through a distance $z$ the effective group velocity of the radiation is $v_g=(dk/d\omega)^{-1}=c - z \omega/\psi$, and (\ref{FreeSpace}) becomes
\begin{eqnarray}
|\phi_{\rm det}(t)| = \frac{\Delta}{\sqrt{2 \pi}} \exp \left[- \frac{(t-z/v_g)^2 \Delta^2}{2} \right].
\end{eqnarray}
Phase perturbations due to gravitational inhomogeneities are not dispersive because the group velocity is independent of frequency.  As a result, there is no additional decorrelation due to any dependence of the effective refractive index of the lensing medium on frequency.

We now investigate the effect of an inhomogeneous mass distribution on the temporal broadening of the signal.  Using the decomposition given by equation (\ref{WaveDecomp}), the time-varying intensity may be written as
\begin{eqnarray}
I_{\rm det}(t) = \langle \phi_{\rm det}(t) \phi_{\rm det}^*(t) \rangle = \frac{1}{(2 \pi)^2} \int_{-\infty}^{\infty} \langle \tilde \phi(k) \tilde \phi^*(k') \rangle \, f_B(\omega) f_B^*(\omega') \, e^{ i z(k - k') - i t( \omega - \omega')} \, d\omega \, d\omega' ,
\end{eqnarray}
where the wavenumbers $k$ and $k'$ are functions of $\omega $ and $\omega'$ respectively.  The terms inside the integral depend, in general, on both the frequency $\omega$ and the frequency difference $\omega-\omega'$.  However, if the bandwidth of the receiver is small, the quantity $\langle \tilde \phi(k,z,{\bf r}) \tilde \phi^*(k',z,{\bf r}) \rangle$ depends strongly only on the frequency difference $\Delta \omega=\omega-\omega'$, and its dependence on the central frequency $\omega$ (or $\omega'$) can be neglected.  One then writes $\Delta \omega=\omega-\omega'$, $W=(\omega+\omega')/2$ and performs the integral over $W$, leaving 
\begin{eqnarray}
I_{\rm det} (t)
&=& \frac{1}{(2 \pi)^2} \int d\Delta\omega\,  \langle \tilde \phi(\omega) \tilde \phi^*(\omega +\Delta \omega) \rangle \left[ \sqrt{\pi} \Delta \exp\left( - \frac{\Delta \omega^2}{4 \Delta^2} \right) \right] e^{i \Delta \omega(z/v_g - t)} .
\end{eqnarray}
This is recognized as a convolution with respect to $t$, so that
\begin{eqnarray}
I_{\rm det}(t)  = P_1 (t) \star P_2(t), \label{Convol}
\end{eqnarray}
with
\begin{eqnarray}
P_1(t) &=& \frac{1}{2 \pi} \int \left[ \sqrt{\pi} \Delta \exp\left( - \frac{\Delta \omega^2}{4 \Delta^2} \right) \right] e^{i \Delta \omega(z/v_g - t)} d\Delta\omega = \Delta^2 \exp \left[ - (t- z/v_g)^2 \Delta^2 \right], \label{P1}
\end{eqnarray}
and
\begin{eqnarray}
P_2(t) 
&=& \frac{c}{2 \pi} \int_{-\infty}^{\infty} d\Delta k \, \langle \tilde \phi(k) \tilde \phi^*(k+\Delta k) \rangle e^{i c \Delta k(t-z/v_g)}. \label{P2}
\end{eqnarray}
In practice the convolution described in (\ref{Convol}) means that either $P_1$ or $P_2$ is primarily responsible for the temporal broadening, depending on which of the two is the broadest function of $t$.  The term $P_1$ represents the temporal intensity variation observed from a pulsed signal after propagating through a homogeneous gravitational potential.  In the present case, when the detector bandpass is large, this term takes the limiting form $\delta (t-z/_g)$, which represents the arrival of a sharp pulse at time $t=z/v_g$.  More generally, when the detector bandpass is large we can write the temporal intensity variations as the convolution of the intensity variations intrinsic to the source, $I_{\rm src}(t)$, with the term $P_2(t)$:
\begin{eqnarray}
I_{\rm det} (t) = I_{\rm src}(t) \star P_2(t).
\end{eqnarray}
The term $P_2$ embodies the effect of mass inhomogeneities on the temporal broadening.  This term is the Fourier transform of the visibility between the wave field measured at some wavenumber $k_1=k$ with the same field measured at wavenumber $k_2=k+\Delta k$, and with ${\bf X}=0$.  Using equation (\ref{Fresnel}), this mean visibility is
\begin{eqnarray}
V(k_1,k_2,z) = \langle \tilde \phi(k_1,z) \tilde \phi^*(k_2,z) \rangle 
&=&   \frac{\left( 1 - \frac{\Delta k^2}{4 k^2} \right) }{(2 \pi r_{\rm F}^2)^2 }  
\left\langle \int d^2 {\bf x}' d^2 {\bf x'}  \exp \left\{ \frac{i}{2 r_{\rm F}^2} 
\left[ x^2 \left( 1-\frac{\Delta k}{2k} \right) - x'^2 \left(1+\frac{\Delta k}{2k} \right) \right]   
\right. \right. \nonumber \\ 
&\null& \left. \left.   \hskip 2in + 
i \psi({\bf x}) \left( 1-\frac{\Delta k}{2k} \right) - i \psi({\bf x}') \left( 1+\frac{\Delta k}{2k} \right) \right\} \right\rangle , \label{MutCoher}
\end{eqnarray}
where we have written $k_1=k+\Delta k/2$ and $k_2 = k-\Delta k/2$.  The quantity $\langle \tilde \phi(k_1,z) \tilde \phi^*(k_2,z) \rangle$ describes the spectral decorrelation of the signal between wavenumbers $k_1$ and $k_2$ caused by its propagation through the inhomogeneous gravitational potential between the source and the observer.  We assume that the phase fluctuations obey gaussian statistics in order to compute the average over the phase fluctuations. However, in interpreting the physical origin of effects due to temporal broadening, we shall see the results obtained here are quite generic, and the assumption of gaussian statistics has little bearing on the nature of the effects.  We also neglect terms of order $\Delta k^2/k^2$ or higher except in front of the potentially large $C_\psi(0)$ term.  The mean visibility evaluates to (see Appendix \ref{MutCoherAppendix}) 
\begin{eqnarray}
V(k_1,k_2,z) &=&  \frac{1 }{(2 \pi)^2 } 
 \int d^2 {\bf r} \, d^2 {\bf q}  \exp \left[- i {\bf r} \cdot {\bf q}   + \frac{i q^2 \Delta k r_{\rm F}^2 }{2 k^2} 
   	- \frac{\Delta k^2 }{k^2} C_\psi(0) - \frac{1}{2} D_\psi \left( {\bf r} \right)  \right] .
\label{MutCoherRes}
\end{eqnarray}
Equation (\ref{MutCoherRes}) is identical in form to the equivalent expression found by Lee \& Jokipii (1975) for scattering broadening in the interstellar medium due to a thin layer of electron density fluctuations, and it is convenient to follow their treatment of the effect of the phase fluctuations on the profile of a temporally broadened signal.  Replacing Eq. (\ref{MutCoherRes}) in Eq. (\ref{P2}), $P_2$ can be represented as the convolution of two broadened profiles
\begin{eqnarray}
P_2(t) = P_R (t) \star P_D(t),
\end{eqnarray}
where 
\begin{eqnarray}
P_R(t) &=& \frac{c}{2 \pi} \int d\Delta k \exp \left[ - i c \Delta k (t - z/v_g) - \frac{\Delta k^2}{k^2} C_\psi(0) \right], \quad \hbox {and} \\
P_D(t)  &=& \frac{c}{(2 \pi)^3} \int d\Delta k \,d^2{\bf q}\, d^2{\bf r} \, \exp \left[ - i c \Delta k (t - z/v_g) 
- i {\bf r} \cdot {\bf q} +  \frac{i\Delta k}{k} q^2 r_{\rm F}^2  - \frac{1}{2} D_\psi \left( {\bf r} \right) \right].
\end{eqnarray}
We refer to $P_R$ and $P_D$ as the refractive and diffractive broadening terms respectively, for reasons explained below.  Performing the integrals over $\Delta k$, one obtains a closed expression for $P_R$ and writes $P_D$ in terms of a Fourier transform:
\begin{eqnarray}
P_R(t) &=& \frac{c k }{2 \sqrt{\pi C_\psi(0)} }  
\left[ - \frac{c^2 k^2  \left( t - \frac{z}{v_g} \right)^2}{4 C_\psi(0)} \right], \quad \hbox {and} \label{PR} \\
P_D(t)  &=& \frac{1}{(2 \pi)^2} \int d^2 {\bf r} d^2{\bf q} \, \delta \left(t -\frac{z}{v_g} - \frac{q^2 r_{\rm F}^2}{2 kc} \right) \, \exp
\left[- i {\bf r} \cdot {\bf q} - \frac{D_\psi({\bf r})}{2}  \right], \\
&=& \frac{1}{(2 \pi)^2} \int d^2 {\bf r} \exp
\left[- i {\bf r} \cdot {\bf q} - \frac{D_\psi({\bf r})}{2}  \right], 
		\qquad \hbox{where } q=\frac{1}{r_{\rm F}}  \sqrt{c k \left( t- \frac{z}{v_g} \right) } . \label{PD}
\end{eqnarray}
The effect on the pulse shape due to these two terms is qualitatively different.  The pulse broadening due to the refractive term is a gaussian symmetric about the mean pulse arrival time with width
\begin{eqnarray}
t_R = \frac{2}{c k} \sqrt{C_\psi (0) }. \label{TRefractive}
\end{eqnarray}
The symmetry of the refractive pulse broadening profile implies that the pulse can arrive either earlier or later than the mean arrival time $t=z/v_g$.  One understands the origin of this effect in terms of deviations in the total gravitational potential along the line of sight.  These give rise to deviations in the total phase delay along the line of sight about the mean value $\langle \psi \rangle$.  The time delay due to a deviation of phase $\delta \psi$ is just $\delta \psi/c k$, so the root mean square phase delay expected from variations in the total gravitational potential is $\langle \delta \psi^2 \rangle^{1/2}/ ck$, which is exactly the temporal broadening time given above.  Refractive broadening only affects the mean pulse profile over a long time (see \S6.1), and is not relevant to the effect depicted in Fig.\,2.  This is because the broadening occurs only on a time scale over which the total phase delay along the ray path varies, which is, in practice, large.

On the other hand, pulse broadening due to the diffractive term is asymmetric: inspection of Eq. (\ref{PD}) shows that $P_D(t)$ is zero before the pulse arrival time $t=z/v_g$.  Unlike refractive broadening, diffractive broadening affects the shape of each individual pulse, and is directly applicable to the effect described in Fig.\,2.  The exact form of the diffractive broadening term has not been determined analytically, except for the specific case $D_\psi = (r/r_{\rm diff})^2$, where one derives 
\begin{eqnarray}
P_D(t) = \left\{ \begin{array}{ll} 
0, & t-z/v_g < 0, \\
\frac{r_{\rm diff}^2}{2 \pi} \exp \left[ - c k (t-z/v_g)  r_{\rm diff}^2/2  r_{\rm F}^2 \right], & t-z/v_g > 0.
\end{array} \right. 
\end{eqnarray}
The temporal broadening time scale is estimated for other phase structure functions by inspection of equation (\ref{PD}).  The diffractive profile begins to decline when $1/q$ is equal to the scale on which the term $\exp[- D_\psi({\bf r})/2]$ varies.  This occurs on the scale $r=r_{\rm diff}$, so $P_D(t)$ declines on the time scale given by $\sqrt{c k (t - z/v_g)}/r_{\rm F} = r_{\rm diff}^{-1}$.  Thus the temporal broadening time is
\begin{eqnarray}
t_D = \frac{1}{c k} \frac{r_{\rm F}^2}{r_{\rm diff}^2}. \label{TDres}
\end{eqnarray}
The origin of the diffractive broadening term lies in the fact that the observer receives radiation from a range of angles over the lensing screen.  Radiation that is scattered through a larger angle takes longer to reach an observer.  The typical scattering angle is $\theta_{\rm scat}= 1/k r_{\rm diff}$.  For an observer located a distance $z$ behind the lensing plane the time delay associated with radiation arriving from angle $1/k r_{\rm diff}$ is $z (c k^2 r_{\rm diff}^2)^{-1}$, which is equivalent to (\ref{TDres}) above.  The geometry associated with this time delay is illustrated in Fig.\,3. Numerical estimates of the scatter broadening profile due to the diffractive term are given in Lee \& Jokipii (1975).

\begin{figure}
\centerline{\psfig{file=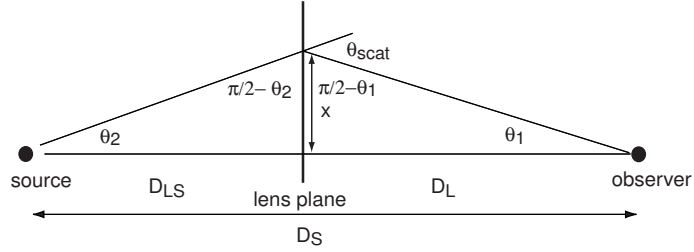,width=9cm}}
\caption{Diffractive temporal broadening arises because radiation, scattered through a typical angle $\theta_{\rm scat} = (k r_{\rm diff})^{-1}$, is delayed relative to radiation propagating directly along the line of sight to the source.}
\end{figure}


\section{Models for the mass fluctuations}  \label{Models}
Since the foregoing results depend on the power spectrum of the surface mass density fluctuations, it is appropriate to quantify the form and magnitude of this spectrum.  We consider the power spectrum of mass fluctuations in a galaxy due to its stellar, gas and dark matter content.  Although it might be supposed that dark matter dominates the mass content of a galaxy, the extent to which it contributes to gravitational scattering relative to other forms of matter depends critically on its distribution.  With the nature of dark matter largely unknown, the power spectrum of dark matter fluctuations on scales relevant to those probed by gravitational scattering is ill-constrained at present.  Nonetheless, it is necessary to consider the conditions under which the effects are likely be measurable, even if the estimates are rough. 

The results derived here are employed in the following section to estimate the magnitude of the effects caused by gravitational scattering.

\subsection{Scattering by a star field} \label{Starfield}
Here we consider the mass power spectrum due to a collection of $N$ stars, each of mass $M$ and with positions ${\bf r}_1, {\bf r}_2, \ldots, {\bf r}_N$.  The mass surface density is written in the form
\begin{eqnarray}
\Sigma({\bf r}) = \sum_i^N M \, f({\bf r} - {\bf r}_i),  \label{MassForm}
\end{eqnarray}
where  $M f({\bf r}) $ is the projected surface mass density of a star centred on the origin.  
We compute the mass surface density covariance:
\begin{eqnarray}
\langle \Sigma ({\bf r}' + {\bf r}) \Sigma ({\bf r}') \rangle &=&  \frac{1}{A} \int d^2{\bf r}' \Sigma ({\bf r}' + {\bf r}) \Sigma ({\bf r}') = \frac{1}{(2 \pi)^2 A} \int d^2{\bf q} e^{-i {\bf q} \cdot {\bf r} } | \tilde \Sigma ({\bf q}) |^2.
\label{StarCorrelation}
\end{eqnarray}
The power spectrum of the surface density is 
\begin{eqnarray}
 |\tilde \Sigma({\bf q}) |^2 = \sum_{i , j}^N M^2 | \tilde f({\bf q}) |^2 e^{i {\bf q} \cdot ({\bf r}_i - {\bf r}_j)},
\end{eqnarray}
and its average is computed by separating the $N$ self ($i=j$) contributions from the $N(N-1)$ cross-term ($i \neq j$) contributions:
\begin{eqnarray}
|\tilde \Sigma({\bf q}) |^2 = N M^2 | \tilde f({\bf q}) |^2 +
N (N-1) M^2 | \tilde f({\bf q}) |^2 \left\langle e^{i {\bf q} \cdot ({\bf r}_i - {\bf r}_j)} \right\rangle . \label{PwrSum}
\end{eqnarray}
The average over positions ${\bf r}_i$ and ${\bf r}_j$ appearing in the last term vanishes if the masses are distributed randomly over all space.  However, this average is in general non-zero and one has
\begin{eqnarray}
\left\langle e^{i {\bf q} \cdot ({\bf r}_i - {\bf r}_j)} \right\rangle = \frac{1}{A^2} \int d^2{\bf r}_i d^2 {\bf r}_j \, p_2({\bf r}_i, {\bf r}_j) e^{i {\bf q} \cdot ({\bf r}_i - {\bf r}_j)} , \label{PosnAvgs}
\end{eqnarray}
where $p_2({\bf r}_i, {\bf r}_j)$  is the joint probability of finding an object at position ${\bf r}_i$ while another is located at position ${\bf r}_j$.  In practice this average is never zero because the idealization of stars distributed randomly on an infinitely extended lensing plane is unrealistic.  The objects are always confined within a finite area, however large, so there is always some maximum separation between pairs ${\bf r}_{\rm max}={\bf r}_i-{\bf r}_j$.  Thus there is always a sufficiently small $q \sim 1/r_{\rm max}$ for which the average in (\ref{PosnAvgs}) is equal to one.  However, in certain situations it is convenient to separate out the large-scale distribution of matter, which may give rise to large phase gradients and hence gravitational macrolensing, from the local variations in mass density due to stars.  In other words, it is often convenient to treat the distribution of stars as being locally uniform, even when the stellar distribution cannot be uniform on large scales.

Now if the distribution of object positions depends strongly on the difference, $\Delta{\bf r} = {\bf r}_i-{\bf r}_j$, and only weakly on the average position ${\bf s} = ({\bf r}_i+{\bf r}_j)/2$, we can approximate the distribution of object positions, as described by the joint probability $p_2$, as being wide-sense stationary.  We write $p_2({\bf r}_i, {\bf r}_j)$ as a function of $\Delta {\bf r}$ only, which gives 
\begin{eqnarray}
\left\langle e^{i {\bf q} \cdot ({\bf r}_i - {\bf r}_j)} \right\rangle &=& \frac{1}{A^2} \int d^2\Delta {\bf r}
 \, d^2 \Delta {\bf s} \, p_2(\Delta {\bf r}) e^{i {\bf q} \cdot \Delta {\bf r}} 
= \frac{\tilde p_2({\bf q})}{A} . \label{PosnAvg}
\end{eqnarray}

Placing these averages back into equation (\ref{StarCorrelation}) and, using the fact that the mean surface density. $\langle \Sigma \rangle$ is $(N/A) M \tilde f({\bf q}=0)$, we obtain the average autocorrelation of the mass surface density fluctuations
\begin{eqnarray}
C_\Sigma ({\bf r}) &=& \frac{\sigma M^2}{(2 \pi)^2} \left[ 
 \int d^2 {\bf q}  e^{-i {\bf q} \cdot {\bf r} } | f({\bf q}) |^2  \left(  1  + \sigma \tilde p_2({\bf q}) \right) \right] - \sigma^2 M^2  \tilde f^2(0),
\end{eqnarray}
where $\sigma = N/A$ is the mean surface density of objects.  The corresponding power spectrum of surface density fluctuations is 
\begin{eqnarray}
\Phi_\Sigma ({\bf q}) &=&  \sigma M^2  | f({\bf q}) |^2 \left[  1 + \sigma \tilde p_2({\bf q}) \right]  - \sigma^2 M^2 \delta ({\bf q}) \tilde f^2(0). \label{PhiStars}
\end{eqnarray}
This power spectrum is inserted directly into equation (\ref{DtoPhi}) to determine the structure function of phase perturbations due to gravitational lensing by a field of stars.

\subsubsection{Scattering by a homogeneous stellar distribution}
We construct the phase structure function due to a population of $N$ objects with mass $M$.  In order to construct a model that is analytically tractable, we idealize the radial density profile of the objects as a gaussians with radius $R$: $\rho(x,y,z) = M\, (2 \pi R^2)^{-3/2} \exp[- (x^2+y^2+z^2)/2 R^2]$.  The surface density of each object is then $\Sigma = M\, (2 \pi R^2)^{-1} \exp[- (x^2+y^2)/2 R^2]$, from which we identify (cf. eq. (\ref{MassForm}))
\begin{eqnarray}
f({\bf r})= \frac{1}{2 \pi R^2} \exp \left[ -\frac{r^2}{2 R^2} \right], \qquad \hbox{and} \qquad \tilde f({\bf q}) = \exp \left[ - \frac{q^2 R^2}{2} \right], \qquad \hbox{where } r=\sqrt{x^2 + y^2}, \quad q=\sqrt{q_x^2+q_y^2}.
\end{eqnarray}

When the outer scale of the stellar distribution is much larger than the scale sizes of the lensing objects, one separates the contribution due to the large density gradient introduced by the overall stellar distribution from the local mass density fluctuations caused by individual stars.  If the stars are distributed homogeneously locally, then the power spectrum of local mass fluctuations reduces to $\Phi_\Sigma ({\bf q}) = \sigma M^2 | \tilde f({\bf q})|^2$, where we neglect the unimportant term proportional to $\delta({\bf q})$ in eq. (\ref{PhiStars}).  The phase structure function, $D_\psi$, is constructed by substituting the power spectrum of mass surface density fluctuations in Eq. (\ref{DtoPhi}):
\begin{eqnarray}
D_\psi ({\bf r}) = 4 \pi K^2 \sigma M^2 \int dq \, q^{-3} \left[1- J_0(q r) \right] \, \exp \left[- q^2 R^2 \right]. 
\end{eqnarray} 
The $q^{-3}$ weighting term of the integrand causes problems with the convergence of the structure function as $q \rightarrow 0$, because the rest of the integrand does not increase sufficiently quickly at small $q$.  It is convenient to separate part of the integrand, $1-{\rm J}_0(q r)$, into two terms $(1- J_0(q r) - q^2 r^2/4)$ and $q^2 r^2/4$.  With this separation, one has 
\begin{eqnarray}
D_\psi ({\bf r}) &=& 4 \pi K^2 \sigma M^2  (I_1 + I_2)  \\
I_1 &=& \int_0^\infty dq \, q^{-3} 
\left (1- J_0(q r) - \frac{q^2 r^2}{4} \right)  e^{-q^2 R^2}  
= \frac{R^2}{2} \left( e^{-r^2/4 R^2} - 1- \Gamma \right) - \frac{r^2 + 4 R^2}{8} 
\left[ E_1 \left( \frac{r^2}{4 R^2} \right) + 
\log \left( \frac{r^2}{4 R^2} \right) \right]  + \frac{(2- \Gamma ) r^2  }{8} , \label{I1}
\\
I_2 &=&  \frac{r^2}{4} \lim_{q' \rightarrow 0} \int_{q'}^\infty dq \, q^{-1} e^{-q^2 R^2}
	= - \frac{r^2}{8} \lim_{q \rightarrow 0}  {\rm Ei} \left( -q^2 R^2 \right) = - \frac{r^2}{8} \lim_{q \rightarrow 0}  \left( \Gamma + \log (q^2 R^2) \right) , \label{I2} 
\end{eqnarray}
The integral $I_1$ converges because the term $(1-J_0(q r) - q^2 r^2/4)$ contains only terms ${\cal O}(q^4)$ and higher.  The integral $I_2$ diverges because it contains terms ${\cal O}(q^{-1})$.  However, as can be seen in eq. (\ref{I2}), the divergence is only logarithmic in $q$, so we approximate the contribution due to $I_2$ as $r^2$ multiplied by a term of order unity.  Physically, this is justified because there must always be some large scale at which the power spectrum of mass density fluctuations cuts off because the distribution of stars is finite.  In practice the stars are only distributed homogeneously over a small volume; in an exact treatment their true distribution on large scales must also be taken into account, as discussed in Katz, Balbus \& Pacyznski (1986).   These authors also identify and describe the origin of this logarithmic divergence in the context of microlensing by stars.

For small $r$, the sum $I_1+I_2$ scales as $r^2$ times a term of order unity (due to the logarithmic cutoff) and the phase structure function is
\begin{eqnarray}
D_\psi(r) \sim 4 \pi K^2 \sigma M^2 r^2 = 1.94 \times 10^{-5} \, (1+z_L)^2 \nu^2 
\left( \frac{M}{1\,{\rm M}_\odot } \right)^2 \left( \frac{\sigma}{100\,{\rm stars \,pc}^{-2}} \right) 
\left( \frac{r}{1 \,{\rm pc}} \right)^2.
\end{eqnarray}
For realistic stellar densities, this implies a very large diffractive scale length,  
\begin{eqnarray}
r_{\rm diff} &=& 2.2 \times 10^2 \, (1+z_L)^{-1} \nu^{-1} \left( \frac{M}{1\,{\rm M}_\odot } \right)^{-1} \left( \frac{\sigma}{100\,{\rm stars \,pc}^{-2}} \right)^{-1/2} \, {\rm pc}.  \label{RdiffHomog}
\end{eqnarray}
We note that $r_{\rm diff}$ scales proportional to $N^{1/2}$, which implies that the typical scattering angle due to lensing by uniformly distributed stars scales as $\theta_{\rm scat} \propto N^{-1/2}$.  This result has also been derived in the context of microlensing of a homogeneous distribution of stars by Katz, Balbus \& Pacyznski (1986).  These authors point out that if the stellar distribution is not uniform on large scales -- as must be the case in practice -- the large scale stellar distribution also causes macrolensing, which shears the microlensed images.  This effect is implicitly ignored here because we are only interested in the stochastic fluctuations in the gravitational phase delay, and not in large scale phase gradients due to macrolensing.

The magnitude of $r_{\rm diff}$ is large, which is a reflection of the fact that, in a uniform distribution, each star's contribution to the power spectrum is independent of all other stars.  The power spectrum of mass density fluctuations scales only linearly with the number density of lensing objects, $\sigma$, rather than as $\sigma^2$, when clustering is important (viz. eq. (\ref{PhiStars}).  Evidently, stellar clustering must be important if stars are to make an appreciable contribution to the power spectrum of mass surface density fluctuations in a galaxy.

\subsection{Scattering by gas} \label{Gas}

Observations of ionized plasma and neutral hydrogen in the local universe suggest that the gas distribution in galaxies follows a power law with an index between $-3$ and $-4$ from sub-parsec to kiloparsec scales (Armstrong, Rickett \& Spangler 1995; Dickey et al. 2001; Stanimirovic \& Lazarian 2001; Braun 1999).  This motivates us to consider a model for the gas whose volume density fluctuations follow a power law power spectrum in between minimum and maximum spatial wavenumbers:
\begin{eqnarray}
\Phi_\rho ({\bf q}) = \left\{ \begin{array}{ll}
0, & q < q_{\rm min} \\
A \,q^{-\beta}, & q_{\rm min} < q < q_{\rm max} \\
0, & q > q_{\rm max},
\end{array} \label{3DGasSpectrum}
\right. 
\end{eqnarray}
where the amplitude $A$ is chosen so that the total variance in the volume density is $\sigma_\rho^2$, and we assume $q_{\rm min} \ll q_{\rm max}$.  For $\beta < 3$ the mass variance is dominated by fluctuations at the inner scale, $l_0=1/q_{\rm max}$, with $A= 4 \pi (3-\beta) \, q_{\rm max}^{\beta-3} \, \sigma_\rho^2 $.  On the other hand, with $\beta > 3$ the mass variance is dominated by fluctuations at the outer scale, $L_0=1/q_{\rm min}$, with $A=4 \pi \, (\beta-3) \, q_{\rm min}^{\beta-3} \, \sigma_\rho^2$.

The two-dimensional power spectrum of surface mass density fluctuations is obtained by projecting the power spectrum of volume mass density fluctuations, $\Phi_\rho$, onto the lensing plane.  For a gravitational wave propagating along the $z$-axis one sets the argument $q_z$ in $\Phi_\rho$ to zero, so the resulting power spectrum of surface mass density fluctuations  is $\Phi_\Sigma(q_x,q_y)= \Delta L \, \Phi_\rho(q_x,q_y,q_z=0)$, where $\Delta L \ga L_0$ is the thickness scattering medium along the $z$-axis.  (One can think of the quantity $\Phi_\rho(q_x,q_y,q_z=0)$ as the power in surface mass density fluctuations per unit length.) 
The phase structure function due to these mass fluctuations is, using (\ref{DtoPhi}),
\begin{eqnarray}
D_\psi ({\bf r}) = 4 \pi \, A\,\Delta L\, K^2 \int_{q_{\rm min}}^{q_{\rm max}} q^{-3-\beta} \left[ 1- J_0(q r) \right]. 
\end{eqnarray}
For spectra steeper than $\beta > -2$ the phase structure function is strongly dominated by mass fluctuations at the outer scale, and one has
\begin{eqnarray}
D_\psi(r) =  \frac{4 \pi A \, \Delta L\,K^2}{2+\beta} q_{\rm min}^{-2-\beta} \left[1- \null_1F_2 \left(-1-\frac{\beta}{2} ; 1, -\frac{\beta}{2} ; - \frac{q_{\rm min}^2 r^2}{4} \right) \right].
\end{eqnarray}
The phase structure function saturates at the outer scale of the distribution, for $r \approx L_0$.
For small arguments, $q_{\rm min}^2 r^2 = r^2/L_0^2 < 1$ and $\beta > 0$ we can expand the hypergeometric function $\null_1F_2$ to yield
\begin{eqnarray}
D_\psi(r) =  \frac{4 \pi^2 (3-\beta) \,\Delta L\, K^2 \sigma_\rho^2}{\beta} r^2 \times \left\{ \begin{array}{ll}
  l_0^2 L_0 \left(\frac{L_0}{l_0} \right)^\beta  , & 0 < \beta < 3 \\
- L_0^3 , & \beta > 3. \\
\end{array}
\right. 
%
\end{eqnarray}
We identify the quantity $M_\sigma^2 \equiv \sigma_\rho^2 L_0^6$ as the total variance in the mass fluctuations encompassed within each cell of size $L_0$, which allows us to express the phase structure function normalized to realistic quantities for an $L_*$ galaxy.  The phase structure function takes the form $D_\psi(r) = (r/r_{\rm diff})^2$ with, for $\beta > 3$, 
\begin{eqnarray}
r_{\rm diff} = 0.018\, \nu^{-1} (1+z_L)^{-1} \, \left( \frac{\beta-3}{\beta} \right)^{-1/2} 
\left( \frac{M_\sigma}{5 \times 10^7\,{\rm M}_\odot} \right)^{-1} \, \left( \frac{\Delta L}{2\,{\rm kpc}} \right)^{-1/2} \, \left( \frac{ L_0}{1\,{\rm kpc}} \right)^{3/2} {\rm pc}. \label{RdiffGas}
\end{eqnarray}
We have normalized the diffractive scale length to a line of sight in which the gas fluctuations extend along a distance of $2 \,$kpc, while the outer scale of the turbulence, $L_0$, is normalized to a value suggested by neutral hydrogen observations of galaxies in the local group.  The mass variance is normalized to a value of only $5 \times 10^7\,$M$_\odot$ at the outer scale of the turbulence.  A typical $L_*$ galaxy contains a mass of $\sim 5 \times 10^9\,$M$_\odot$.  If most of this gas is distributed inhomogeneously and contained within a typical volume $\sim 100\,$kpc$^3$, then the variance of mass fluctuations within each cell of size $L_0^3$ is $M_\sigma^2 = (5 \times 10^7\,{\rm M}_\odot)^2$.

The result for $\beta>3$ given in eq. (\ref{RdiffGas}) is the most useful since observations suggest a power law index for gas fluctuations in the range $3 < \beta < 4$.  However, for completeness we include the results for $0 < \beta < 3$, for which the phase structure function is
\begin{eqnarray}
D_\psi (r) =0.0030 \, \times 10^{3 \beta}  \, \left( \frac{3-\beta}{\beta} \right) (1+z_L)^2 \, \nu^2 \, 
 \left( \frac{M_\sigma}{5 \times 10^7\,{\rm M}_\odot} \right)^2 \,  \left(\frac{\Delta L}{2\,{\rm kpc}} \right)
 \left( \frac{ L_0}{1\,{\rm kpc}} \right)^{\beta-5} \, \left( \frac{ l_0}{1\,{\rm pc}} \right)^{2-\beta}
\left( \frac{r}{1\,{\rm pc}} \right)^2 .
 \end{eqnarray}
For instance, the case $\beta=2$ then yields a diffractive scale length of 
\begin{eqnarray}
r_{\rm diff} = 0.026\, \nu^{-1} (1+z_L)^{-1} \, \left( \frac{M_\sigma}{5 \times 10^7\,{\rm M}_\odot} \right)^{-1} \,   \left( \frac{\Delta L }{2\,{\rm kpc}} \right)^{-1/2}
 \left( \frac{ L_0}{1\,{\rm kpc}} \right)^{3/2}  
\,{\rm pc}.
\end{eqnarray}
As the galaxy would in general possess some inclination to the line of sight, the path length through the gas may vary considerably from the value of $\Delta L =2\,$ kpc used in the normalizations here.  

It may be wondered why fluctuations in the gas density of a galaxy, which only account for $\sim 5\times 10^9$\,M$_\odot$ of the mass content of an $L_*$ galaxy, yield a much greater contribution to the phase structure function relative to the large mass $\sim 10^{11}\,$M$_\odot$ associated with stars.  The difference arises because of the assumptions underlying the distributions of the two forms of matter.  The stellar content of the galaxy is assumed to be uniformly distributed and thus explicitly ignores stellar clustering.  The assumption of uniformity ensures that the power spectrum of mass surface density fluctuations increases only linearly with the stellar density.  On the other hand, observations show that gas tends to follow a power law distribution of density fluctuations, indicating that the gas is clustered in a hierarchy of scales.   In this case the power spectrum of surface density fluctuations scales with the square of the gas density.  We note that the presence of stellar clustering in a population of stars would also cause the power spectrum to rise with the square of the stellar density, as demonstrated by equation (\ref{PhiStars}).

\subsection{Dark matter fluctuations}
Here we consider the distribution of dark matter associated with the halo of a galaxy.  Dark matter is also distributed on larger scales (i.e. it is associated with clusters of galaxies), but we concentrate only on dark matter on galactic scales.  Irrespective of the dark matter distribution on larger scales, any extragalactic source of gravitational radiation is guaranteed to at least propagate through the dark matter distribution of its own galaxy in order to reach an observer near Earth. 
Both of the foregoing treatments of mass fluctuations in \S\S \ref{Starfield} \& \ref{Gas} are readily generalized to treat fluctuations in the dark matter, whether it is clumped in star-like objects, or clustered on a hierarchy of scales and thus distributed according to a power law.  


We consider the dark matter associated with a galaxy similar to the Milky Way.  Here the dark matter must extend to at least several kiloparsecs above and below the Galactic plane, and extend out to $\sim 30\,h^{-1}\,$kpc along the plane.  The mass to light integrated over the entire Galaxy is $\sim 30\, {\rm M}_{\odot}/{\rm L}_{\odot}$; however, this ratio increases sharply near the edge of the luminous disk to $\sim 1000 \,{\rm M}_{\odot}/{\rm L}_{\odot}$ (see, e.g., Sofue \& Rubin 2001).

A plausible model expected on the basis of cold dark matter (CDM) models is a power law spectrum of mass fluctuations between some outer scale $L_0$ and some inner scale $l_0$ with an index $\beta \approx 3$ for the range of (galaxy-size) scales of interest here (e.g. Peacock 1999 and references therein).  (The power spectrum is expected to follow a $\beta=3$ index for scales below the horizon at the epoch of matter-radiation equality $\sim 16 (h \Omega)^{-1}\,$Mpc, where $h$ is the Hubble constant in units of 100 km\,s$^{-1}$Mpc$^{-1}$.) 

The model presented in \S\ref{Gas} for gas fluctuations is readily adapted to the present situation.
As for fluctuations in the gas distribution, the two dimensional power spectrum of surface mass density fluctuations is obtained by projecting the power spectrum of volume mass density fluctuations onto the lensing plane.  One then has $\Phi_\Sigma(q_x,q_y)= \Delta L \, \Phi_\rho(q_x,q_y,q_z=0)$, where $\Delta L$ is the thickness scattering medium along the direction of propagation, assumed to be along the $z$-axis.

The power law scattering for $\beta \ga 3$ is independent of the inner scale, and yields
\begin{eqnarray}
r_{\rm diff} = 1.8 \times 10^{-4} \, \nu^{-1} (1+z_L)^{-1} \, \left( \frac{\beta-3}{\beta} \right)^{-1/2} 
\left( \frac{M_\sigma}{10^{11}\,{\rm M}_\odot} \right)^{-1} \,  
\left( \frac{\Delta L}{5\,{\rm kpc}}\right)^{-1/2} 
\,\left( \frac{ L_0}{10\,{\rm kpc}} \right)^{3/2} {\rm pc}. \label{RdiffDM}
\end{eqnarray}
We have normalized the diffractive scale length to a total mass $\sim 10^{11} M_\odot$ distributed inhomogeneously in each cell with radius comparable to the outer scale $L_{\rm out} \sim 10 \,$kpc.  The outer scale may be larger and encompass a greater mass, but if $M_\sigma$ increases roughly proportional to $L_0^3$, the diffractive scale remains constant.  Gravitational scattering effects in the context of this model are expected to be very important, as demonstrated in the following section.

Another possibility, though somewhat implausible, is that the dark matter is homogeneously distributed in clumps of uniform size.  In this case the results of \S\ref{Starfield} are applicable.  However, even for dark matter masses $\sim 10^{13}\,{\rm M}_\odot$, the power associated with a uniform distribution of lensing objects is small, and we see from equation (\ref{RdiffHomog}) that the diffractive scale length is many orders of magnitude larger than the Fresnel scale. Clearly, gravitational scattering effects would be unimportant for lensing caused by a uniform distribution of dark matter.

\section{Discussion: Implications for the detection of gravitational waves}

We have developed a general theory describing the variability and temporal broadening of gravitationally scattered gravitational radiation, and have considered models for the mass fluctuations likely to drive these effects.  However, we also need to consider the geometry of the lensing situation in order to  discuss the implications of gravitational scattering on the detectability of gravitational radiation.  A large variety of scattering parameters and geometries is expected, depending on the type of gravitational wave source under consideration and the astrophysical environment in which it is likely to be encountered.  Our aim in this paper is to illustrate that scattering effects are important in many instances likely to be encountered in practice; an exhaustive treatment of all the circumstances under which gravitational scattering may be important is beyond the scope of the present paper.  

Gravitational radiation always encounters at least two regions of mass fluctuations along its path toward Earth.  It must propagate through the star, gas and the halo of dark matter distributed throughout the source's host galaxy and, to be detected at Earth, it must also propagate through the matter distribution associated with our own Galaxy.  Propagation through the matter distribution of the host galaxy is particularly important for certain sources of gravitational radiation, such as coalescing supermassive black hole binaries, as these objects are likely to be located at the centres of galaxies.  The radiation is then subject to gravitational scattering through a line of sight intersecting the densest environments of the host galaxy.  Which of the three constituents of a galaxy  -- gas, stars or dark matter -- contributes most to the scattering depends on the distribution of the matter in each of these three forms.

Gravitational radiation may encounter additional mass fluctuations due to any intervening galaxy interposed along the line of sight between the source and the detector.  Although this eventuality seldom occurs for sources of gravitational radiation in the nearby universe, it is worth considering in the context of lensing of powerful sources located at high redshifts, where a line of sight is likely to intersect several galaxies and proto-galaxies.  For example, the sensitivity of LISA is expected to be sufficient to detect coalescence of supermassive binary black holes out to redshifts $z \sim 20$, and in this case the radiation would likely propagate through a large number of intervening systems.

The geometry of lensing of an extragalactic source due to matter in the Milky Way is similar to lensing by the host galaxy, for reasons discussed below.  For a source located in the plane of our own Galaxy, the results are quantitatively similar to the lensing of an extragalactic source by its host galaxy.

\subsection{The time scale of wave amplitude decorrelation}

The wave field of a lensed gravitational wave varies due to the movement of fluctuations in the gravitational potential transverse to line of sight.  The time scale of wave amplitude variations is important, because decorrelation must occur on a time scale longer than $\nu^{-1}$ for gravitational radiation of frequency $\nu$ to be detected.  The results of \S\ref{MeanVis} show that the wave amplitude fluctuations decorrelate on a time scale $t=r_{\rm diff}/v_{\rm eff}$, which is the time over which the mean square phase on the lensing plane changes by one radian as the phase fluctuations are advected past an observer at speed $v_{\rm eff}$.  This time scale is independent of the distance to the mass fluctuations driving the phase fluctuations.

The velocity of lensing material across the line of sight to the source of gravitational waves is uncertain.  The effective speed at which the mass fluctuations move across the line of sight depends on the velocities of the lensing medium, the Earth and the source.  The motion of the source relative to the lensing medium is likely to dominate the effective lensing velocity.  
Typical stellar motions and galactic gas velocities are of the order of one hundred kilometers per second within a galaxy.  If such a speed is representative of motions of most matter within the galaxy, then it is also representative of the expected speed of any source of gravitational waves relative to other constituents of the galaxy.  Thus the typical lensing speed is of order $v_{\rm eff}=100\,$km\,s$^{-1}$.  Lensing of a gravitational wave source by other material in the source's host galaxy gives rise to wave amplitude decorrelation on a time scale
\begin{eqnarray}
t = 1.8 \times 10^3 \, \left( \frac{r_{\rm diff} }{0.001 \,{\rm pc}} \right) \left(\frac{v_{\rm eff}}{100\,{\rm km\,s}^{-1}} \right)^{-1}\, \hbox{days} .
\end{eqnarray}
Temporal decorrelation of the wave amplitude is expected to occur much more rapidly at higher frequencies.  The diffractive scale is inversely proportional to frequency in the lensing models considered in the previous section.  Higher relative velocities are expected if one considers lensing by an external galaxy that intersects the line of sight to the source, where the peculiar speed of a galaxy transverse to the line of sight is of order 1000\,km\,s$^{-1}$. 

Scattering-induced temporal decorrelation of the wave field occurs on a time scale much longer than the period of the radiation itself.  Thus we conclude that while scattering may alter the amplitude of the lensed radiation, the decorrelation occurs sufficiently slowly that the wave field is well correlated from one wave period to the next.

\subsection{Temporal broadening}

Temporal broadening asymmetrically smears the intensity profile of any intrinsically variable  gravitational wave source with broadening time $t_D =  r_{\rm F}^2/r_{\rm diff}^2 \omega$.  Temporal broadening is important when the ratio $r_{\rm F} / r_{\rm diff}$ is large, and it depends critically on the Fresnel scale, and thus on the lensing geometry. The Fresnel scale depends on the ratio of lensing angular diameter distances $D_{\rm eff} \equiv D_L D_{LS}/D_S$ (viz. eq. (\ref{rF})) which is well approximated by the source-lens distance when the lensing occurs in the host galaxy (i.e. $D_{LS} \approx D_S$), and by the observer-lens distance when the lensing occurs in our Galaxy.   In a Euclidean spacetime (in which $D_S=D_L+D_{LS}$), the maximum value of $D_{\rm eff}$ occurs when the lensing plane is midway between the source and observer, and $D_{\rm eff}=D_L/2$.  However, in our universe the angular diameter distance saturates at $z \sim 1$ and falls slowly for higher redshifts. Thus, for large distances between the source, lensing plane and observer, the slow decline of angular diameter distances beyond $z \sim 1$ implies that the maximum effective angular diameter distance is well approximated by its saturation value, $D_{\rm eff} \sim 1\,$Gpc.

The thin screen approximation used here is an excellent approximation when treating lensing by an intervening galaxy because the thickness of the lensing material is small compared to the effective distance, $D_{\rm eff}$ to the lensing material.  However, the validity of this approximation is less obvious when treating lensing due to mass fluctuations in the host galaxy of a source, where the fluctuations occur over a range of distances that is potentially large relative to $D_{\rm eff}$.  However, the results derived here are still applicable provided that one choses an effective distance $D_{\rm eff}$ that is characteristic of the distance to the bulk of the lensing material.   For instance, in treating the temporal broadening scattering of pulsar radiation due to interstellar scintillation, in which the interstellar medium is extended along the entire path from the source to the observer, one takes $D_{\rm eff}$ to be of order half the distance to the pulsar.  Moreover, a rigorous generalization of the results for an extended medium shows that the scatter broadening time may still be written as $D_{\rm eff} \theta_c^2/2c$, where $\theta_c=(k r_{\rm diff})^{-1}$ is the characteristic scattering angle (Lee \& Jokipii 1975).  This argument is applicable to the present situation, as the physics underlying interstellar scintillation and gravitational-induced temporal smearing is identical.

In terms of normalized values typical of scattering in the host galaxy and an intervening galaxy, the Fresnel scale for lensing at a distance $D_{\rm eff}$ is 
\begin{eqnarray}
r_{\rm F} &=& 0.0028 \,  \nu^{-1/2} (1+z_L)^{-1/2} \, \left(\frac{D}{5\,{\rm kpc}} \right)^{1/2} \, {\rm pc} 
= 1.2 \, \nu^{-1/2} (1+z_L)^{-1/2} \, \left(\frac{D}{1\,{\rm Gpc}} \right)^{1/2} \, {\rm pc}.
\end{eqnarray}
The large value of the Fresnel scale typical for matter located in between the host galaxy and the Milky Way enhances the effectiveness of temporal smearing.

First consider temporal broadening due to lensing caused by the mass distribution of the host galaxy.   The diffractive scale estimated for lensing due a uniform distribution of stars is so large that temporal smearing due to scattering by stars is unimportant.  The effect of scattering by the gas content of a galaxy depends on the index of its power law distribution.  Observations suggest an index $\beta=3-4$ (e.g. Dickey et al. 2001), so the diffractive scale is of order $0.018 \, \nu^{-1} (1+z_L)^{-1}\,$pc for the gas content typical of an $L_*$ galaxy.  Thus lensing due to gas density fluctuations alone causes temporal smearing on time scales of order $4\,$ms. 

Scattering effects due to the large mass associated with dark matter fluctuations are potentially important.  For instance, the diffractive scale estimated for lensing by a power law of dark matter fluctuations with an index $\beta=3.2$ is $r_{\rm diff} \approx 7.2 \times 10^{-4} \, \nu^{-1} (1+z_L)^{-1}$.  With a Fresnel scale $r_{\rm F}=0.0028 \,\nu^{-1/2} (1+z_L)^{-1/2}$\,pc, the temporal broadening time is $t_D \approx 2\,$s.  Temporal broadening on this time scale would severely decrease the detectability of sources whose intrinsic variations occur on a comparable time scale, and render undetectable any intrinsic variations substantially shorter than this scatter broadening time.  The temporal broadening time is independent of frequency for the models of the matter distribution considered in \S\ref{Models}.  

The large value of the Fresnel scale associated with lensing by a galaxy interposed along the line of sight renders the temporal intensity variability of most sources of gravitational radiation undetectable.  The broadening time for the dark matter diffractive scale considered above and the Fresnel scale $1.2 \,\nu^{-1/2} (1+z_L)^{-1/2} \,$pc is $t_D \approx  4 \times 10^5\,$s.  The broadening time due to gas fluctuations alone is $t_D \approx 7 \times 10^2\,$s.

The statistical approach adopted in the present treatment of temporal broadening is only valid when a large number of lensing lumps of matter -- be they stars, gas or dark matter -- contribute to the lensing at any instant.  This is guaranteed to be the case if a large number of objects are contained within a volume of radius $r_{\rm F}$ integrated along the line of sight.  However, as the Fresnel radius scales as $\nu^{-1/2}$, there must be a sufficiently high transition frequency at which the statistical approach becomes invalid.   Thus, although the diffractive broadening time is independent of frequency, there must be a sufficiently high frequency at which the character of the temporal smearing changes.
When relatively few objects contribute to the lensing, there is only a small number of time delays associated with the small number of ray directions that scatter radiation into an observer's line of sight.   An observer would then receive a small number of delayed copies of the same intrinsic intensity variations.  This transition frequency depends strongly on the minimum scale length on which matter exhibits structure.  If it is distributed inhomogeneously on scales smaller than the Fresnel scale then the present statistical treatment is always valid.  The statistical approach is an excellent approximation when considering scattering by an intervening galaxy, where the Fresnel scale is particularly large.

The statistical treatment of lensing in the host galaxy is reasonable if the dark matter is cold.  For dark matter particles of rest mass energy $E_m$, the spectrum of fluctuations is expected to be cut off at the free-streaming scale (e.g. Padmanabhan 1993),
\begin{eqnarray}
\lambda_{\rm fs} = 0.005 \left( \frac{E_m}{1\,{\rm GeV}c^{-2}} \right)^{-4/3} \,{\rm pc},  
\end{eqnarray}
which, for cold ($E_m> 1\,$GeV$c^{-2}$) dark matter particles, is comparable to or below the scales typically probed by gravitational scattering in a host galaxy for wave frequencies $\nu \ga 1$\,Hz.  On the other hand, dark matter particles considerably lighter than this are expected to exhibit little structure on scales probed by gravitational lensing; a statistical approach would then no longer be appropriate, and the character of the scattering would be qualitatively different.  In this sense, measurements of temporal broadening over a range of frequencies, and thus length scales, provide an exquisitely sensitive means of measuring the nature of dark matter.

\section{Conclusions}

The gravitational potentials of mass fluctuations encountered by gravitational radiation as it propagates toward Earth perturb its wave front, causing an observer to perceive variability in the wave amplitude and temporal smearing of the signal.  We employ a statistical approach to relate these effects to the underlying power spectrum of mass density perturbations that drive the scattering.  This approach is motivated by the fact that gravitational radiation produced by extragalactic sources is likely to encounter the gravitational potentials of many objects along its path toward Earth.

Mass fluctuations along the line of sight to a source of gravitational radiation induce variations in its wave amplitude as it propagates toward Earth.  The characteristic fluctuation time scale is the time on which the wave amplitude fluctuations decorrelate.   This occurs over the interval over which the root mean square phase difference along the line of sight to the source changes by one radian.  
This time scale is frequency dependent because the phase induced by gravitational perturbations is linearly proportional to frequency.  Wave amplitude variability therefore occurs more rapidly at high frequencies. We consider lensing by gas and dark matter, and estimate that decorrelation occurs on a time scale $\ga 10^3 \nu^{-1}\,$days.  However, even when temporal variations are too slow to be discerned, the random amplification of the radiation due to focusing and defocusing of the radiation still alters the detectability of the signal.  

The most important scattering effect relates to the temporal smearing of scattered radiation, as it has a direct bearing on the detectability of any source whose intensity, the square of the wave amplitude, varies with time.  The lightcurve of a temporally smeared source in a stochastic medium is the intrinsic intensity lightcurve of the source convolved with a function whose width is the temporal broadening time scale, $t_D$.  Temporal smearing arises in a scattering medium because radiation scattered through an angle and deflected back towards towards the line of sight takes longer to reach an observer than radiation propagating directly along the line of sight.  The magnitude of this effect depends on the power spectrum of the mass fluctuations and on the lensing geometry.  The power spectrum depends on the particular distribution and density of stars, gas and dark matter along the line of sight.  Gravitational radiation is always subject to scattering, as it must escape through the mass distribution of its host galaxy and propagate through the mass distribution of the Milky Way in order to be detected.


We consider the consequences of lensing due to a power law distribution of mass fluctuations, due to either gas or dark matter; we also consider a simple model due to lensing by a collection of stars.  
These estimates suggest that temporal smearing due to the host galaxy is dominated by the dark matter distribution, and is important for radiation whose intensity varies on a time scale less than a second.  The temporal smearing time is independent of the observing frequency.  Temporal smearing effects scale linearly with the effective distance to the scattering material, $D_{\rm eff}=D_L D_{LS}/D_S$, so the smearing time due to lensing by an external galaxy is exceptionally large.  For lensing at cosmological distances, $D_{\rm eff} \sim 1\,$Gpc, the temporal smearing time is estimated to be of order $t_D\sim 10^5\,$s.  As the distribution of dark matter on small scales relevant to temporal broadening is ill-constrained, we caution that the foregoing estimates are highly uncertain. 
When dark matter is absent, even the gas content of a galaxy can appreciably broaden the intrinsic intensity variations of a source.  The broadening time associated with the gas content of the host galaxy is of order milliseconds, while the time associated with scattering due an intervening system is of order $10^3$\,s.

The circumstances in which temporal smearing poses a serious limitation on the detectability of intrinsic source variations depends on the distribution of scattering material.  The description of temporal smearing presented here is valid when one is justified in treating the phase fluctuations using a statistical approach.  However, it fails when only a few inhomogeneities contribute to the scattering at any one instant, at which point the observed intensity lightcurve becomes the sum of a small number of delayed copies of the intrinsic intensity lightcurve.  Thus its validity varies according to the line of sight and the scale on which matter distributed along it is inhomogeneous.  Clearly, the scale on which dark matter is inhomogeneous is critically important.  If it is inhomogeneous on scales smaller than the Fresnel scale, as is expected for CDM fluctuations, then the statistical approach adopted for temporal broadening here is well justified.  Thus, the scale on which dark matter is distributed is of critical importance in determining the detectability of gravitational radiation.  Conversely, temporal smearing measurements of gravitational radiation place strong constraints on the distribution and hence the nature of dark matter along any line of sight through which a source of gravitational radiation is detected.




\begin{acknowledgements}
The author thanks Don Melrose, Mark Walker and Ole M\"oller for their encouragement and comments, and Marco Spaans for suggesting additional improvements to the manuscript.
\end{acknowledgements}

\appendix

\section{Derivation of mutual coherence function $V(k_1,k_2;{\bf r}=0)$} \label{MutCoherAppendix}
Here we calculate the mutual coherence function, $\langle \tilde \phi(k+\Delta k/2,z,{\bf r}) \tilde \phi^*(k-\Delta k/2.z,{\bf r}) \rangle$, following on from equation (\ref{MutCoher}).  We perform the average over phase fluctuations assuming that $\psi$ is a normally distributed random variable with zero mean (the mean phase delay is incorporated into the term involving the group velocity $v_g$ in equation (\ref{P1})):  
\begin{eqnarray}
V(k_1,k_2,z) 
&=&  \frac{\left( 1 - \frac{\Delta k^2}{4 k^2} \right) }{(2 \pi r_{\rm F}^2)^2 } 
\left\langle \int d^2 {\bf x}' d^2 {\bf x'}  \exp \left\{ \frac{i}{2 r_{\rm F}^2} 
\left[ x^2 \left( 1-\frac{\Delta k}{2k} \right) - x'^2 \left(1+\frac{\Delta k}{2k} \right) \right]   
- \frac{\Delta k^2 }{k^2} C_\psi(0) - 
\frac{1}{2} D_\psi({\bf x}-{\bf x}') \left( 1 - \frac{\Delta k^2}{k^2} \right)  \right\} \right\rangle . \end{eqnarray}
With the change of variables 
\begin{eqnarray}
{\bf r} &=& {\bf x} \left( 1-\frac{\Delta k}{2k} \right)^{1/2} 
		- {\bf x}' \left( 1+\frac{\Delta k}{2k} \right)^{1/2}, \nonumber \\
{\bf s} &=& \frac{1}{2} \left[ {\bf x} \left( 1-\frac{\Delta k}{2k} \right)^{1/2} 
	+ {\bf x}' \left( 1+\frac{\Delta k}{2k} \right)^{1/2} \right],
\end{eqnarray}
which has a Jacobian $1+ {\cal O}(\Delta k^2 /k^2)$ which we approximate as unity,  the mutual coherence becomes
\begin{eqnarray}
V(k_1,k_2,z) &=&  \frac{\left( 1 - \frac{\Delta k^2}{4 k^2} \right) }{(2 \pi r_{\rm F}^2)^2 } 
\left\langle \int d^2 {\bf r} d^2 {\bf s}  \exp \left\{ \frac{i {\bf r} \cdot {\bf s} }{r_{\rm F}^2} 
   - \frac{\Delta k^2 }{k^2} C_\psi(0) - 
\frac{1}{2} D_\psi \left( {\bf r} + \frac{\Delta k}{2k} {\bf s} \right) \left( 1 - \frac{\Delta k^2}{k^2} \right)  \right\} \right\rangle . 
\end{eqnarray}
We make the further change of variables ${\bf r}' = {\bf r} + {\bf s} \Delta k/ 2k$ and ${\bf s}'={\bf s}$, which again has a Jacobian $1+{\cal O}(\Delta k^2/k^2)$, to write
\begin{eqnarray}
V(k_1,k_2,z) &=&  \frac{\left( 1 - \frac{\Delta k^2}{4 k^2} \right) }{(2 \pi r_{\rm F}^2)^2 } 
 \int d^2 {\bf r}' d^2 {\bf s}'  \exp \left\{ \frac{i {\bf r}' \cdot {\bf s}' }{r_{\rm F}^2} 
- i \frac{s'^2 \Delta k}{2 r_{\rm F}^2 k^2} 
   - \frac{\Delta k^2 }{k^2} C_\psi(0) - 
\frac{1}{2} D_\psi \left( {\bf r}' \right) \left( 1 - \frac{\Delta k^2}{k^2} \right)  \right\} . 
\end{eqnarray}
A further change of variable ${\bf q}={\bf s}'/r_{\rm F}^2$ yields the final result,
\begin{eqnarray}
V(k_1,k_2,z) &=&  \frac{\left( 1 - \frac{\Delta k^2}{4 k^2} \right) }{(2 \pi)^2 } 
 \int d^2 {\bf r}' d^2 {\bf q}  \exp \left\{ i {\bf r}' \cdot {\bf q}  
- i \frac{q^2 \Delta k r_{\rm F}^2 }{2 k^2} 
   - \frac{\Delta k^2 }{k^2} C_\psi(0) - 
\frac{1}{2} D_\psi \left( {\bf r}' \right) \left( 1 - \frac{\Delta k^2}{k^2} \right)  \right\} . 
\end{eqnarray}
This expression is equivalent to that obtained by Lee \& Jokipii (1975) for interstellar scattering through a medium of plasma inhomogeneities.  

\section{The effective velocity of scattering material across the line of sight} \label{ScatSpeed}

\begin{figure}[h] \label{VelGeom}
\begin{center}
\centerline{\psfig{file=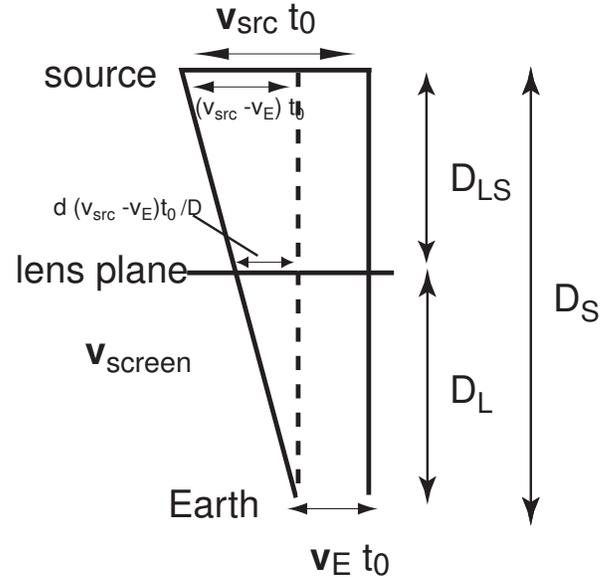,width=180mm}}   
\caption{The geometry showing the distance traversed across the lensing plane in a duration $t_0$.}
\end{center}
\end{figure}

We derive an expression of the velocity of scattering material across the line of sight, taking into account motion in the source, lensing medium, and the Earth.  Consider an object with angular diameter distance $D_S$ from Earth moving at a transverse velocity ${\bf v}_{\rm src}$ measured relative to some frame.  
Between two time intervals $t=0$ and $t=t_0$ the object traverses a
displacement ${\bf v}_{\rm src} t_0$, and the Earth's change in position
is ${\bf v}_{\rm Earth} t_0$, where ${\bf v}_{\rm Earth}$ is measured relative to the same co-ordinate frame.  From Fig.\,\ref{VelGeom} we see that the line of sight moves a distance $\Delta {\bf s}={\bf v}_{\rm Earth} t_0 + ({\bf v}_{\rm src}-{\bf v}_{\rm Earth}) t_0 (D_L/D_S)$ across the scattering screen.  
The effective velocity of the line of sight across the scattering screen
is thus $\Delta {\bf s}/t_0$: 
\begin{eqnarray} 
{\bf v}_{\rm eff} = {\bf v}_{\rm Earth} \left(1 -\frac{D_L}{D_S} \right) + {\bf v}_{\rm src} \left(
\frac{D_L}{D_S} \right) 
\end{eqnarray} 
In this time interval the screen also moves an additional distance $v_{\rm screen} t_0$, so the effective velocity at which the line of sight to the object crosses a point on the screen is
\begin{eqnarray}
{\bf v}_{\rm eff} = {\bf v}_{\rm screen} - \left[ {\bf v}_{\rm Earth} \left(1 -\frac{D_L}{D_S} \right) + {\bf v}_{\rm src} \left( \frac{D_S}{D_L} \right) \right].
\end{eqnarray}

The observed scintillation time scale is the time taken by the line of sight to traverse a length
scale $r_{\rm diff}$ on the screen:
\begin{eqnarray}
t_{\rm scint} = \frac{r_{\rm diff}}{v_{\rm eff}}.
\end{eqnarray}
This is the time scale of the scintillations as observed at Earth.

There is a subtle distinction between the
velocity at which the source-observer line of sight moves across the lensing plane 
and the actual velocity at which an observer moves through the lensing pattern, as it is projected at Earth.  The distinction arises because the wavefronts emitted from the source are spherical, and an observer located near the lensing plane would not measure the same lensing pattern scale length  as an observer
on Earth.  A wavefront ``imprinted'' with a pattern of scale length $r_{\rm diff,screen}$ on the scattering screen has a pattern scale of length $r_{\rm diff,Earth}= r_{\rm diff,screen} D_S/D_{LS}$ upon reaching
Earth.  Thus, for an observer who measures $r_{\rm diff,Earth}$ directly from the scale size of the scintillation pattern passing across the Earth, the real speed of the scintillation pattern at Earth is
\begin{eqnarray}
{ v}_{\rm real} 
&=& \frac{r_{\rm diff,Earth}}{t_{\rm scint} }.
\end{eqnarray}
Thus one has
\begin{eqnarray}
{\bf v}_{\rm real} &=& \frac{\left(\frac{D_S}{D_{LS}}\right) 
r_{\rm diff,screen}}{\frac{r_{\rm diff,screen} }{{\bf v}_{\rm eff}}} 
= \frac{D_S}{D_{LS}} {\bf v}_{\rm screen} - \frac{D_S-D_L}{D_{LS}}{\bf v}_{\rm Earth} - \frac{D_L}{D_{LS}} {\bf v}_{\rm src}.
\end{eqnarray}
This correction is only a concern if we were to measure the scale size of
the lensing pattern directly with an interferometer on Earth.  The expressions derived above 
reduce to those derived in Appendix C of Gupta, Rickett \& Lyne (1994) for scattering in Euclidean space.



\end{document}